\documentclass[amsthm]{elsart}
\usepackage{yjsco}
\usepackage{natbib}
\usepackage{amsmath}
\usepackage{amsfonts}
\usepackage{amssymb}
\usepackage{enumerate}
\usepackage{mathrsfs}
\usepackage[plainpages=true,pdfpagelabels,colorlinks=true,citecolor=blue,hypertexnames=false]{hyperref}

\def\vv {\mathbf{v}}

\def\cD {\mathcal{D}}

\def\bF {\mathbb{F}}

\def\bK {\mathbb{K}}
\def\bL {\mathbb{L}}

\def\bN {\mathbb{N}}

\def\bV {\mathbb{V}}
\def\bW {\mathbb{W}}

\def\bZ {\mathbb{Z}}

\def\si {\sigma}

\def\bigO {\mathcal{O}}

\def\pa{\partial}

\newcommand{\mb} { {\bf B}}

\def\pf{\noindent {\it Proof.\ }}
\def\qed{\hfill \rule{4pt}{7pt}}

\def\si{\sigma}
\def\la{\langle}
\def\ra{\rangle}

\def\pa{{\partial}}

\newtheorem{theorem}{Theorem}[section]
\newtheorem{proposition}[theorem]{Proposition}

\newtheorem{lemma}[theorem]{Lemma}

\newtheorem{remark}[theorem]{Remark}
\newtheorem{define}[theorem]{Definition}
\newtheorem{example}[theorem]{Example}
\newtheorem{problem}[theorem]{Problem}

\renewcommand{\caption}[1]{%
\refstepcounter{theorem}
\centerline{{{\bf Table~\arabic{section}.\arabic{theorem}.} #1} \label{#1}}
}

\begin{document}

\begin{frontmatter}

\title{On the Existence of Telescopers for Rational Functions in Three Variables}

%
%
%
%
%

%
\author{Shaoshi Chen$^{a,b}$, Lixin Du$^{a,b,c}$, Rong-Hua Wang$^{d}$, Chaochao Zhu$^{a,b}$ }
\ead{schen@amss.ac.cn, dulixin17@mails.ucas.ac.cn, wangronghua@tiangong.edu.cn, chaochaozhu@139.com}

\medskip

\address{$^a$KLMM, Academy of Mathematics and Systems Science, Chinese Academy of Sciences,\\ Beijing, 100190, China}
\address{$^b$School of Mathematical Sciences, University of Chinese Academy of Sciences,\\ Beijing, 100049, China }
\address{$^c$Institute for Algebra, Johannes Kepler University, \\ Linz, A4040, Austria}
\address{$^d$School of Mathematical Sciences, Tiangong University, \\Tianjin, 300387, China}


\begin{abstract}
Zeilberger's method of creative telescoping is crucial for the computer-generated proofs of combinatorial and special-function identities.
Telescopers are linear differential or ($q$-)recurrence operators computed by algorithms for creative telescoping. For a given class of inputs,
when telescopers exist and how to construct telescopers efficiently if they exist are two fundamental problems related to creative telescoping.
In this paper, we solve the existence problem of telescopers for rational functions in three variables including 18 cases. We
reduce the existence problem from the trivariate case to the bivariate case and some related problems.
The existence criteria given in this paper enable us to determine the termination of algorithms for creative telescoping with trivariate rational inputs.
\end{abstract}


\end{frontmatter}

\section{Introduction}\label{SECT:intro}
Creative telescoping plays a crucial role in the algorithmic proof theory
of combinatorial identities developed by Wilf and Zeilberger in the early
1990s~\cite{Zeilberger1990, Zeilberger1991, Wilf1992}.
For a given function $f(x, y_1, \ldots, y_n)$, the process of creative telescoping
constructs a nonzero linear differential or ($q$-)recurrence operator $L$ in $x$ such that
\[L(f) = {\Theta_{y_1}(g_1)  + \cdots + \Theta_{y_n}(g_n)},\]
where $\Theta_{y_i}$ denotes the derivation or ($q$-)difference operator in $y_i$
and the $g_i$'s belong to the same class of functions as $f$. The operator $L$ is then called
a \emph{telescoper} for~$f$, and the $g_i$'s
are called the \emph{certificates} of~$L$. Two fundamental problems have been studied extensively
related to creative telescoping.
The first problem is the \emph{existence problem of telescopers}, i.e., deciding the existence of
telescopers for a given class of functions.
The second one is the \emph{construction problem of telescopers}, i.e., designing efficient
algorithms for computing telescopers if they exist. For more open problems related to creative telescoping, one can see~\cite{chen2017JSSC}.
In this paper, we will mainly focus on the existence problem of telescopers and study the construction problem
of telescopers in future work.

The existence of telescopers is closely connected to the termination of algorithms for creative telescoping and the hypertranscendence and algebraic dependency
of functions defined by indefinite sums or integrals~\cite{Hardouin2008, Schneider2010}. In 1990,
Zeilberger first presented a sufficient condition on the
existence of telescopers by showing that telescopers always exist for so-called~\emph{holonomic functions}
in~\cite{Zeilberger1990} using Bernstein's theory of algebraic D-modules.
Soon after this work, Wilf and Zeilberger in~\cite{Wilf1992} proved that telescopers exist for
proper hypergeometric terms. However, holonomicity and properness
are only sufficient conditions. Abramov and Le~\cite{AbramovLe2002} gave a
necessary and sufficient condition on the existence of telescopers for
rational functions in two discrete variables. This work was soon extended to the
hypergeometric case by Abramov~\cite{Abramov2003}, the $q$-hypergeometric case
in~\cite{ChenHouMu2005}, and the mixed rational and hypergeometric
case in~\cite{ChenSinger2012, Chen2015}.
All of the above work only focussed on the problem for
bivariate functions of a special class.
The first criterion on the existence of telescopers beyond the bivariate case was
given in~\cite{Chen2016}, in which a necessary and sufficient condition is presented on the existence problem of telescopers for
rational functions in three discrete variables. The goal of this paper is continuing this project
by considering the remaining cases, in which the continuous, discrete and $q$-discrete variables can appear.

The remainder of this paper is organized as follows. We define the existence problem of telescopers
precisely in Section~\ref{SECT:prelim} and recall different types of reductions that are used in testing
the exactness of bivariate rational functions in Section~\ref{SECT:red}. Existence criteria are given
for 18 types of telescopers for rational functions in three variables in Section~\ref{SECT:criteria}.

A preliminary version~\cite{chen2019ISSAC} of this article has appeared in the Proceedings of ISSAC'19.
In the present version, we include twelve more cases in which the $q$-shift operator appears
and also more detailed proofs throughout.

\medskip
\noindent {\it Acknowledgement.}
The authors would like to thank Ruyong Feng, Hui Huang and Ziming Li for many helpful discussions.
In this work, Shaoshi Chen and Chaochao Zhu was supported by the NSFC grants (No. 11871067 and No. 11688101) and by the Fund of the Youth Innovation Promotion Association, CAS.
Lixin Du was supported by the NSFC grant (No. 11871067) and the Austrian FWF grant (No. P31571-N32). Rong-Hua Wang was supported by the Natural Science Foundation of Tianjin (No. 19JCQNJC14500) and the NSFC grant (No. 11871067).

\section{Preliminaries}\label{SECT:prelim}

Let $\bK$ be a field of characteristic zero and $\bK(\vv)$ be the field of rational functions in the
variables $\vv= \{x, y_1, \ldots, y_n\}$ over $\bK$. For each $v\in \vv$,
the \emph{derivation} $\delta_{v}$ on $\bK(\vv)$ is defined as the usual partial derivation $\partial/\partial_v$
with respect to $v$ satisfying that
$\delta_v(f + g) = \delta_v(f) + \delta_v(g)$ and $\delta_v(fg) = g\delta_v(f) + f \delta_v(g)$
for all $f, g\in \bK(\vv)$.
Moreover, $\delta_v(c)=0$ if and only if
$c\in \bK(\vv\setminus\{v\})$, i.e., $c$ is free of $v$.
For each $v\in \vv$, the \emph{shift operator} $\sigma_v$
is the $\bK$-automorphism of $\bK(\vv)$ defined by $\si_v(v) = v+1$
and $\si_v(w) = w$ for all $w\in \vv\setminus\{v\}$. Let $q\in \bK\setminus\{0\}$ be such that $q^m\neq 1$ for all nonzero $m\in \bZ$.
For each $v\in \vv$,
the \emph{$q$-shift operator} $\tau_{q, v}$ is the $\bK$-automorphism defined by $\tau_{q, v}(v) = qv$
and $\tau_{q, v}(w) = w$ for all $w\in \vv\setminus\{v\}$.
Abusing notation, we let $\delta_v$ and $\theta_v$ with $\theta_v \in \{\si_v, \tau_{q, v}\}$ denote arbitrary extensions
of $\delta_v$ and $\theta_v$ to derivation
and $\overline\bK$-automorphism of $\overline{\bK(\vv)}$, the algebraic closure of $\bK(\vv)$.

Over the field $\bK(\vv)$,  we have a noncommutative algebra
$\cD := \bK(\vv)\langle \partial_x,  \partial_{y_1}, \ldots, \partial_{y_n}\rangle$
in which commutation rules are $\partial_{v_i} \partial_{v_j} = \partial_{v_j} \partial_{v_i}$ for all $v_i, v_j\in \vv$,
and for any $v\in \vv$ and $f\in \bK(\vv)$,
\begin{equation}\label{DEF:delta}
\partial_v f  = \left\{ \begin{array}{ll}
                          f\partial_v + \delta_v(f) & \mbox{if $\partial_v= D_v$,} \\
                          \si_v(f)\partial_v & \mbox{if $\partial_v = S_v$,} \\
                           \tau_{q, v}(f) \partial_v & \mbox{if $\partial_v = T_{q, v}$.} \\
                          \end{array} \right.
\end{equation}
where $D_v, S_v,$ and~$T_{q, v}$ refer to the differential, shift and $q$-shift operators, respectively.
The algebra~$\cD$ is also called the ring of linear functional operators or Ore
polynomials (for more details, see~\cite{BronsteinPetkovsek1996, ChyzakSalvy1998}). Let
$\Delta_v$ be the difference operator $S_v-1$ and $\Delta_{q, v}$ be the $q$-difference operator $T_{q, v}-1$.
For each $v\in \vv$, we define
\begin{equation}\label{DEF:Theta}
\Theta_v := \pa_v - \pa_v(1) = \left\{ \begin{array}{ll}
                          D_v & \mbox{if $\partial_v = D_v$,} \\
                          \Delta_v & \mbox{if $\partial_v= S_v$,} \\
                          \Delta_{q, v} & \mbox{if $\partial_v = T_{q, v}$.} \\
                          \end{array} \right.
\end{equation}
The action of the operator $\partial_v \in \cD$ on an element $f\in \bK(\vv)$ is defined as
\begin{equation}\label{DEF:action}
\partial_v(f)  = \left\{ \begin{array}{ll}
                          \delta_v(f) & \mbox{if $\partial_v= D_v$,} \\
                          \si_v(f) & \mbox{if $\partial_v = S_v$,} \\
                           \tau_{q, v}(f) & \mbox{if $\partial_v = T_{q, v}$.} \\
                          \end{array} \right.
\end{equation}
In general, the action of the operator $L= \sum_{i_0, i_1, \ldots, i_n\geq 0} a_{i_0,  i_1, \ldots, i_n} \partial_x^{i_0} \partial_{y_1}^{i_1}\cdots \partial_{y_n}^{i_n}
\in \cD$ on $f \in \bK(\vv)$ is defined as
\[L(f)= \sum_{i_0, i_1, \ldots, i_n\geq 0} a_{i_0, i_1, \ldots, i_n} \partial_{x}^{i_0}\partial_{y_1}^{i_1}\cdots \partial_{y_n}^{i_n}(f).\]
Then the field $\bK(\vv)$ becomes a left $\cD$-module. In this paper, we will mainly work with rational functions in three variables $x, y, z$
and the operators in $\bK(x, y, z)\langle \partial_x, \partial_y, \partial_{z}\rangle$.
\begin{example}\label{EXAM:action}
Let $L = 1+ (x+yz)D_x + 2S_yT_{q, z}  \in \bK(x, y, z)\langle D_x, S_y, T_{q, z}\rangle$ and $f = 1/(x+yz)$. Then we have
\[L\cdot f =f +(x+yz)\delta_x(f)+ 2\si_y(\tau_{q, z}(f)) = \frac{2}{qz+qyz+x}. \]
\end{example}
The functions we consider will be in certain $\cD$-module,
such as the field $\bK(\vv)$ or $\overline{\bK(\vv)}$.
The ring $\bK(x)\langle \partial_x\rangle$ is a subring of $\cD$ that is also
a left Euclidean domain. Efficient algorithms for
basic operations in $\bK(x)\langle \partial_x\rangle$, such as computing
the least common left multiple (LCLM) of operators, have
been developed in~\cite{BronsteinPetkovsek1996, AbramovLeLi2005}.
\begin{lemma}\label{LEM:lclm}
For an operator $L = \sum_{i=0}^\rho e_i D_x^i\in \overline{\bK(x)}\la D_x \ra$ with $e_{\rho} =1$,
we let $\bF$ be a finite normal extension of $\bK(x)$ containing the coefficients $e_i$'s
and $G$ be the Galois group of $\bF$ over~${\bK(x)}$. Let $T$ be the LCLM of the operators $\si(L) = \sum_{i=0}^\rho \si(e_i)D_x^i$
for all $\si\in G$. Then $T$ belongs to $\bK(x)\la D_x\ra$.
\end{lemma}
\begin{proof}
It suffices to show that $\tau(T) = T$ for all $\tau\in  G$.
Since $D_x$ commutes with any automorphism in $G$~by~\cite[Theorem 3.2.4~(i)]{BronsteinBook}, we have $\tau(L_1 L_2) = \tau(L_1)\tau(L_2)$
for all $L_1, L_2\in \bF\la D_x \ra$.
For each $\si \in G$, we have $T = P_{\si} \si(L)$ for some $P_\si \in \bF\la D_x \ra$, which implies that
$\tau(\si(L))$ divides $\tau(T)$. When $\si$ runs through all elements of $G$, so does $\tau\si$.
Hence $\tau(T)$ is also a common left multiple of the operators $\si(L)$
for all $\si\in G$. Since $\tau(T)$ and $T$ are both monic and of the same degree in $D_x$,
we get $\tau(T)=T$.
\end{proof}
\begin{rem}
The above assertion is not true in the ($q$-)shift case. For example, take $L = S_x + \sqrt{x}$.
The LCLM of $L$ and its conjugation $S_x -\sqrt{x}$ is $S_x^2 - \sqrt{x(x+1)}$, which is not in $\bK(x)\la S_x\ra$.
\end{rem}

\begin{define}\label{DEF:telescoper}
For a rational function~$f\in \bK(x, y, z)$, a nonzero
operator~$L(x, \pa_x)\in \bK(x)\langle \pa_x \rangle$ is called a \emph{telescoper of type\/~$(\pa_x, \Theta_y, \Theta_z)$} for~$f$
if there exist rational functions $g, h\in \bK(x, y, z)$ such that
\begin{equation}\label{EQ:telescoper}
L(x, \pa_x)(f) = \Theta_y(g) + \Theta_z(h).
\end{equation}
The rational functions $g, h$ are called the \emph{certificates} of $L$.
\end{define}

Note that all of the telescopers for a given function together with the zero operator form a left
ideal of $\bK(x)\langle \partial_x \rangle$ (see~\cite[Definition 1]{Chyzak2009}).
The following lemma summarizes closure properties related to the existence of telescopers.
\begin{lemma} \label{LM:closure}
Let $f, g \in \overline{\bK(x, y, z)}$, $a, b \in \bK(x)$ and $\alpha, \beta\in \overline{\bK(x)}$. Then we have
\begin{itemize}
\item[(i)] if both $f$ and $g$ have telescopers in $\bK(x)\la D_x\ra $ of type $(D_x,\Theta_y,\Theta_z)$, so does $\alpha f + \beta g$;
\item[(ii)] if both $f$ and $g$ have telescopers in $\bK(x)\la \pa_x \ra$ of type $(\pa_x,\Theta_y,\Theta_z)$ with $\pa_x\in\{S_x,T_{q,x}\}$, so does $a f + b g$.
\end{itemize}
\end{lemma}
\begin{proof}
We first show that $\alpha f$ has a telescoper in $\bK(x)\la D_x\ra$ if $f$ does.
When $\alpha=0$ the conclusion is obvious.
Next we assume that $\alpha\neq 0$ and
$L = \sum_{i=0}^\rho e_i D_x^i \in \bK(x)\langle D_x\rangle$ is a telescoper for $f$.
Then
$L(f) = \Theta_{y}(u) +  \Theta_{z}(v)$ with
$u, v\in \overline{\bK(x, y, z)}$.
Set $\tilde{L} = L \cdot \frac{1}{\alpha}$, which belongs to $\overline{\bK(x)}\langle D_x\rangle$.
Then we have $\tilde{L}(\alpha f) = \Theta_{y}(u) +  \Theta_{z}(v)$, which means $\tilde{L}$
is a telescoper for $\alpha f$. By Lemma~\ref{LEM:lclm}, there exists
$T\in \bK(x)\la D_x\ra$ such that $T$ is a left multiple of $\tilde{L}$. So $T$ is also a telescoper for $\alpha f$.
When telescopers are in $\bK(x)\la S_x \ra$ or $\bK(x)\la T_{q, x} \ra$, the above argument works for $af$ for any $a\in \bK(x)$.
It remains to show that $f+g$ has a telescoper in $\bK(x)\la \partial_x\ra$ with $\partial_x \in \{D_x, S_x, T_{q, x}\}$ if both $f$ and $g$ do. Assume that $P, Q\in \bK(x)\la \partial_x\ra$
are telescopers for $f, g$, respectively. Then the LCLM of $P$ and $Q$
is a telescoper for $f+g$ by the commutativity between operators in $\bK(x)\la \partial_x\ra$ and the operators $\Theta_{y}$ and $\Theta_z$.
\end{proof}

\begin{center}
\begin{table}
\begin{center}
\begin{tabular}{|c|c|c|}
\hline
Classes & Types &  Telescoping equations \\    \hline
1.  & \text{1.1.}\ $(D_x, D_y,  D_z) $        &
$\begin{array}{c}
L(x, D_x)(f) = D_y(g) + D_z(h).
\end{array}$
\\   \hline
2. & $\begin{array}{l}
\text{2.1.}\  (D_x, \Delta_y, \Delta_z) \\
\text{2.2.}\  (D_x, \Delta_{q, y}, \Delta_{z}) \\
\text{2.3.}\   (D_x, \Delta_{q, y}, \Delta_{q, z})
\end{array}$        &
$\begin{array}{c}
 L(x, D_x)(f) = \Delta_y(g) + \Delta_z(h)  \\
L(x, D_x)(f) = \Delta_{q, y}(g) + \Delta_{z}(h)\\
 L(x, D_x)(f) = \Delta_{q, y}(g) + \Delta_{q, z}(h)
\end{array}$
\\   \hline
3. & $\begin{array}{l}
\text{3.1.}\  (S_x, D_y, D_z) \\
\text{3.2.}\  (T_{q, x}, D_y, D_z)
\end{array}$        &
$\begin{array}{c}
 L(x, S_x)(f) = D_y(g) + D_z(h)  \\
L(x, T_{q, x})(f) = D_y(g) + D_z(h)
\end{array}$
\\   \hline
4. & $\begin{array}{l}
\text{4.1.}\  (S_x, \Delta_y, D_z) \\
\text{4.2.}\  (S_x, \Delta_{q, y}, D_z) \\
\text{4.3.}\   (T_{q, x}, \Delta_y, D_z) \\
\text{4.4.}\   (T_{q, x}, \Delta_{q, y}, D_z)
\end{array}$        &
$\begin{array}{c}
 L(x, S_x)(f) = \Delta_y(g) + D_z(h)  \\
L(x, S_x)(f) = \Delta_{q, y}(g) + D_z(h) \\
 L(x, T_{q, x})(f) = \Delta_y(g)+ D_z(h) \\
 L(x, T_{q, x})(f) = \Delta_{q, y}(g) + D_z(h)
\end{array}$
\\   \hline
5. & $\begin{array}{l}
\text{5.1.}\  (S_x, \Delta_y, \Delta_z) \\
\text{5.2.}\  (S_x, \Delta_{q, y}, \Delta_{z}) \\
\text{5.3.}\   (S_x, \Delta_{q, y}, \Delta_{q, z}) \\
\text{5.4.}\   (T_{q, x}, \Delta_y, \Delta_z) \\
\text{5.5.}\   (T_{q, x}, \Delta_{q, y}, \Delta_{z}) \\
\text{5.6.}\   (T_{q, x}, \Delta_{q, y}, \Delta_{q, z})
\end{array}$        &
$\begin{array}{c}
 L(x, S_x)(f) = \Delta_y(g) + \Delta_z(h)  \\
 L(x, S_x)(f) = \Delta_{q, y}(g) + \Delta_{z}(h) \\
 L(x, S_x)(f) = \Delta_{q, y}(g) + \Delta_{q, z}(h) \\
 L(x, T_{q, x})(f) = \Delta_y(g) + \Delta_z(h)\\
 L(x, T_{q, x})(f) = \Delta_{q, y}(g) + \Delta_{z}(h) \\
 L(x, T_{q, x})(f) = \Delta_{q, y}(g) + \Delta_{q, z}(h)
\end{array}$
\\   \hline
6. & $\begin{array}{l}
\text{6.1.}\  (D_x, \Delta_y, D_z) \\
\text{6.2.}\ (D_x, \Delta_{q, y}, D_z)
\end{array}$       &
$\begin{array}{c}
 L(x, D_x)(f) = \Delta_y(g) + D_z(h)  \\
 L(x, D_x)(f) = \Delta_{q, y}(g) + D_z(h)
\end{array}$
\\   \hline
\end{tabular}

\medskip
\caption{Six different classes of existence problems of telescopers} \label{Table:types}
\end{center}
\end{table}

\end{center}

Let $V = (V_1, \ldots, V_m)$ be any set partition of the variables $\vv = \{x, y_1, \ldots, y_n\}$.
A rational function $f\in \bK(\vv)$ is said to be \emph{split} with respect to the partition $V$
if $f = f_1\cdots f_m$ with $f_i\in \bK(V_i)$. A polynomial $p\in \bK[\vv]$ is said to be \emph{integer-linear} in $\bK[\vv]$
if there exist $r\in \bK[z]$ and $a, b_1, \ldots, b_n\in \bZ$ such that $p = r(ax + b_1y_1 + \cdots + b_ny_n)$.
A polynomial $p\in \bK[\vv]$ is said to be \emph{$q$-integer-linear} in $\bK[\vv]$
if there exist $r\in \bK[z]$ and $a, b_1, \ldots, b_n, s, t_1, \ldots, t_n\in \bZ$ such that $p = x^sy_1^{t_1}\cdots y_n^{t_n} r(x^ay_1^{b_1}\cdots y_n^{b_n})$,
A rational function $f=P/Q\in \bK(\vv)$ with $P, Q\in \bK[\vv]$ and $\gcd(P, Q)=1$ is said to be \emph{($q$-)proper} in $\bK(\vv)$ if
$Q$ is a product of ($q$-)integer-linear polynomials over $\bK$.  Split polynomials and ($q$-)proper rational
functions will be used to state our existence criteria for telescopers in Section~\ref{SECT:criteria}.

In the subsequent sections, we will study the existence of telescopers for rational functions in three variables.
More precisely, we consider the following problem.

\medskip \noindent
{\bf Existence Problem for Telescopers.}\,\,
For a rational function~$f\in \bK(x, y, z)$, decide the existence
of telescopers of type~$(\pa_x, \Theta_y, \Theta_z)$ for~$f$.

\begin{remark} \label{RE:types}
In the trivariate case, there are 18 different types of telescopers up to the symmetry among $(\Theta_y, \Theta_z)$ which
are collected into six different classes in Table~\ref{Table:types} according to different techniques used in the studies.
\end{remark}

Different types of partial fraction decompositions will be
used in solving the existence problems of telescopers.
Let~$G = \langle \theta_{x},\theta_y, \theta_z\rangle$ be the free abelian group generated by the
operators~$\theta_{x},\theta_{y}, \theta_{z}$ with $\theta_v \in \{\si_v, \tau_{q, v}\}$.
Let~$f\in \bK(x, y, z)$ and $H$ be a subgroup of~$G$.
We call the set
\[ [f]_H := \{c\cdot \psi(f)\mid \text{$\psi\in H$ and $c\in \bK\setminus\{0\}$}\}\]
the \emph{$H$-orbit} at~$f$. Two elements~$f, g\in \bK(x, y, z)$ are said to be \emph{$H$-equivalent} if~$[f]_H = [g]_H$, denoted by $f \sim_H g$. The relation
$\sim_H$ is an equivalence relation in $\bK(x, y, z)$. Let $f = P/Q$ and $g = A/B$ with $P, Q, A, B\in \bK[x, y, z]$,
$\gcd(P, Q)=1$ and $\gcd(A, B)=1$.
If $f \sim_H g$, then $P\sim_H Q$ and $A\sim_H B$ since any $\psi\in H$ is an automorphism on $\bK(x, y, z)$.
So detecting the $H$-equivalence among rational functions
can be reduced to that among polynomials. Two irreducible polynomials in distinct $H$-orbits are clearly coprime.
A nonzero rational function $f\in \bK(x, y, z)$ is said to be~\emph{$(\theta_x, \theta_y, \theta_z)$-invariant} if there exist $m, n, k\in \bZ$, not all zero, and $c\in \bK\setminus\{0\}$ such that $\theta_{x}^m\theta_{y}^n\theta_{z}^k(f) = c\cdot f$. By comparing
the leading coefficients, the constant $c$ in the above relation must be of the form $q^s$ for some $s\in \bZ$.
Moreover,  $c=1$ if all $\theta_{x},\theta_{y}$, and $\theta_{z}$
are shift operators.



For any subgroup $H$ of $G$ and any polynomial $Q\in \bK(x, y)[z]$, one can group all of irreducible factors in $z$ of $Q$
into distinct $H$-orbits that leads to the factorization
\[Q = c \cdot \prod_{i=1}^n \prod_{j=1}^{m_i} \psi_{i, j}(d_i)^{e_{i, j}}, \quad \text{where $c\in \bK(x, y), n, m_i, e_{i, j}\in \bN$ and $\psi_{i, j}\in H$}\]
and the $d_i$'s are monic irreducible polynomials in distinct $H$-orbits. With respect to this factorization, we have the unique partial
fraction decomposition for a rational function $f = P/Q\in \bK(x, y, z)$ of the form
\begin{equation}\label{EQ:hdecomp}
f = p + \sum_{i=1}^n\sum_{j=1}^{m_i}\sum_{\ell=0}^{e_{i, j}}\frac{a_{i, j, \ell}}{\psi_{i, j}(d_i)^\ell},
\end{equation}
where $p, a_{i, j, \ell}\in \bK(x, y)[z]$ satisfying that $\deg_{z}(a_{i, j, \ell}) < \deg_z(d_i)$.
In the sequel, we will take different $H$ according to different types of existence problems.
\begin{example}\label{EXAM:hdecomp}
Consider the rational function of the form
\[f = {\frac {x}{{z}^{2}+2x+y}}+{\frac {y}{{z}^{2}+2x+y+1}}+{\frac {-yz+
x}{{z}^{2}+2qx+y}}+\,{\frac {3{x}^{2}}{{z}^{2}+2qx+y+2z+2}}.
\]
If $H = \la \si_y \ra$, then we have the decomposition
\[f = \frac{x}{d_1} + \frac{y}{\si_y(d_1)} + \frac{-yz+x}{d_2} + \frac{3x^2}{d_3},\]
where $d_1 = z^2 + 2x+y$, $d_2 = {z}^{2}+2qx+y$ and $d_3 = {z}^{2}+2qx+y+2z+2$. Note that $d_1, d_2, d_3$
are in distinct $\la \si_y \ra$-orbits. If $H = \la \tau_{q, x}, \si_y \ra$,  then we have the different decomposition
\[f = \frac{x}{d_1} + \frac{y}{\si_y(d_1)} + \frac{-yz+x}{\tau_{q, x}(d_1)} + \frac{3x^2}{d_3},\]
where $d_1, d_3$ are in distinct $\la \tau_{q, x}, \si_y \ra$-orbits. If $H = \la \tau_{q, x}, \si_y, \si_z \ra$, then we have
another decomposition
\[f = \frac{x}{d_1} + \frac{y}{\si_y(d_1)} + \frac{-yz+x}{\tau_{q, x}(d_1)} + \frac{3x^2}{\tau_{q, x}\si_y\si_z(d_1)}.\]

\end{example}

\section{Reductions and Exactness Criteria}\label{SECT:red}

In this section, let $\bF$ be any field of characteristic zero and will take $\bF = \bK(x)$ in Section~\ref{SECT:criteria}.
In order to detect the existence of telescopers, we first need to check whether 1 is a telescoper or not. This is equivalent to the following problem.

\medskip \noindent
{\bf Exactness Testing Problem.}
For a rational function~$f\in \bF(y, z)$, decide whether there exist $g, h\in \bF(y, z)$ such that
\[f = \Theta_y(g) + \Theta_z(h).\]
If such $g, h$ exist, we say that $f$ is \emph{$(\Theta_y, \Theta_z)$-exact} in $\bF(y, z)$.

\begin{remark} \label{RE:exact}
Since there are three choices for each operator in $\{\Theta_y, \Theta_z\}$
together with the symmetry between $\Theta_y$ and $\Theta_z$, there are 6 different types of exactness testing problems, listed in Table~\ref{Table:exact}.
\begin{center}
\begin{table}
\begin{center}
\begin{tabular}{|c|c|}
\hline
Cases & Exactness equations  \\    \hline
Continuous case & \text{1.1.}\ $f = D_y(g) + D_z(h)$ \\ \hline
Discrete cases & $\begin{array}{l}
\text{2.1.}\  f = \Delta_y(g)+ \Delta_z(h) \\
\text{2.2.}\  f =  \Delta_{q,y}(g)+\Delta_z(h)  \\
\text{2.3.}\   f = \Delta_{q, y}(g)+ \Delta_{q, z}(h)
\end{array}$ \\ \hline
Mixed cases &  $\begin{array}{l}
\text{3.1.}\  f = \Delta_y(g)+ D_z(h) \\
\text{3.2.}\  f =  \Delta_{q,y}(g)+D_z(h)
\end{array}$
\\   \hline
\end{tabular}

\medskip
\caption{Six different cases of exactness testing problems } \label{Table:exact}
\end{center}
\end{table}

\end{center}

\end{remark}

The following lemma shows that the exactness is unchanged even when we are looking for the $g$ and $h$
in a larger field.

\begin{lemma}\label{LEM:trace}
Let $f\in \bF(y, z)$.
Then $f$ is $(\Theta_y,\Theta_z)$-exact in $\overline{\bF(y, z)}$
if and only if it is $(\Theta_y,\Theta_z)$-exact in $\bF(y, z)$.
\end{lemma}
\begin{proof}
The sufficiency is obvious. For the necessity, we assume that there exist
$u, v \in \overline{\bF(y, z)}$ such that $f = \Theta_{y}(u) + \Theta_{z}(v)$.
Let $\bL$ be a finite normal extension of $\bF(y, z)$ containing the $u, v$ and $\Theta_{y}(u), \Theta_{z}(v)$ and
let $\text{Tr}_{\bL/\bF(y, z)}$ be the trace from $\bL$ to $\bF(y, z)$, which commutes with ($q$-)shift operators by~\cite[Lemma 3.1]{ChenSinger2014}
and also with derivations  by~\cite[Theorem 3.2.4~(i)]{BronsteinBook}. Then
\[
  \text{Tr}_{\bL/\bF(y, z)}(f)  {=} \text{Tr}_{\bL/\bF(y, z)}\left(\Theta_{y}(u) + \Theta_{z}(v)\right) {=}
  \Theta_y(\text{Tr}_{\bL/\bF(y, z)}(u))+ \Theta_z(\text{Tr}_{\bL/\bF(y, z)}(v)).
\]
Since $f\in \bF(y, z)$, we have $\text{Tr}_{\bL/\bF(y, z)}(f) = m f$ with $m {=} [\bL:\bF(y, z)]$.
Thus $f = \Theta_{y}(g) + \Theta_{z}(h)$ with $g = \frac{1}{m}\text{Tr}_{\bL/\bF(y ,z)}(u)$ and
$h = \frac{1}{m}\text{Tr}_{\bL/\bF(y ,z)}(v)$ that are both in~$\bF(y, z)$.
\end{proof}

Let $\mathcal{E}$ denote the set of all $(\Theta_y, \Theta_z)$-exact rational functions in $\bF(y, z)$. Note that $\mathcal{E}$
forms a subspace of $\bF(y, z)$ viewed as an $\bF$-vector space.
Reduction algorithms have been developed in~\cite{Chen2012, ChenSinger2014, HouWang2015, Chen2018, Wang2020}
for simplifying rational functions modulo $\mathcal{E}$ and then reducing the exactness problem from general rational functions to
simple fractions. For later use, we summarize these reductions as follows.

\subsection{The continuous case}\label{SSUBSECT:differenal}
For a rational function $f\in \bF(y, z)$, the Ostrogradsky--Hermite reduction~\cite{Ostrogradsky1845, Hermite1872} with respect to $z$
decomposes $f$ into the form
\begin{equation}\label{EQ:reduceOH}
  f = D_z(g) + \frac{a}{b},
\end{equation}
where $g\in \bF(y, z)$ and $a, b\in \bF(y)[z]$ with $\gcd(a, b)=1$, $\deg_z(a)< \deg_z(b)$
and $b$ being squarefree in $z$ over $\bF(y)$. Moreover, $f = D_z(u)$ for some $u\in \bF(y, z)$
if and only if $a=0$.  We recall the criterion on the $(D_y, D_z)$-exactness of bivariate rational functions from~\cite[Lemma 4]{Chen2012}.
\begin{lemma}\label{LEM:DDExact}
Let $f\in \bF(y, z)$ be of the form~\eqref{EQ:reduceOH} and write
\[ \frac{a}{b} = \sum_{i=1}^n \frac{\alpha_i}{z-\beta_i},\]
where $\alpha_i, \beta_i\in \overline{\bF(y)}$ with $\beta_i \neq \beta_j$ for $i,j$ with $1\leq i, j \leq n$ and $i \neq j$.
Then $f$ is $(D_y, D_z)$-exact in $\bF(y, z)$ if and only if for each $i$
with $1\leq i \leq n$, we have $\alpha_i = D_y(\gamma_i)$ for some $\gamma_i\in \overline{\bF(y)}$.
\end{lemma}
The above lemma reduces the exactness problem in the differential case from bivariate rational functions
to univariate algebraic functions. Let $\alpha\in \overline{\bF(y)}$ be an algebraic function over $\bF(y)$
with $n := [\bF(y, \alpha):\bF(y)]$. If $\alpha = D_y(\beta)$ for some $\beta \in \overline{\bF(y)}$, then
$\beta \in \bF(y, \alpha)$ by the trace argument as in the proof of Lemma~\ref{LEM:trace}. Assume that
$\beta = b_0 + b_1 \alpha + \cdots + b_{n-1}\alpha^{n-1}$ with $b_i \in \bF(y)$. Then the equality
$\alpha = D_y(\beta)$ leads to a system of linear differential equations on the $b_i$'s, whose rational solutions
can be computed by the method in~\cite{Barkatou1999}. A generalization of the Ostrogradsky--Hermite reduction
to the algebraic case also solves the exactness problem of algebraic functions~\cite{chen2016ISSACb}.

\subsection{The discrete cases}\label{SSUBSECT:shift}
For any automorphism~$\theta$ on $\bF(y, z)$ and~$a, b\in \bF(y, z)$, we have
the reduction formula
\begin{equation}\label{eq:red}
\frac{a}{\theta^n(b)} = \theta(g) - g + \frac{\theta^{-n}(a)}{b},
\end{equation}
where $g = \sum_{i=0}^{n-1} \frac{\theta^{i-n}(a)}{\theta^i(b)}$ if $n\geq 0$ and $g = -\sum_{i=0}^{-n-1} \frac{\theta^{i}(a)}{\theta^{n+i}(b)}$ if $n<0$.
By using the above reduction formula, Abramov's reduction in $z$~\cite{Abramov1975, Abramov1995b} decomposes $f\in \bF(y, z)$ into the form
\begin{equation}\label{EQ:reduceA}
  f = \Delta_z(g) + \frac{a}{b},
\end{equation}
where $g\in \bF(y, z)$ and $a, b\in \bF(y)[z]$ with $\gcd(a, b)=1$, $\deg_z(a)<\deg_z(b)$
and $b$ being shift-free in $z$ over $\bF(y)$, i.e., for any $k\in\bZ\setminus\{0\}$ we have $\gcd(b,\sigma_z^k(b))=1$. Moreover, $f = \Delta_z(u)$ for some $u\in \bF(y, z)$
if and only if $a=0$. We use the reduction formula~\eqref{eq:red} with $\theta= \si_y$ to further decompose $f$ as
\begin{equation}\label{EQ:reduceSS}
  f = \Delta_y(u) + \Delta_z(v) + \sum_{i=1}^I\sum_{j=1}^{J_i} \frac{a_{i, j}}{d_i^j},
\end{equation}
where $u, v\in \bF(y, z)$, $a_{i, j}\in \bF(y)[z]$, and $d_i\in \bF[y, z]$ are such that
$\deg_z(a_{i, j})<\deg_z(d_i)$ and the $d_i$'s are irreducible  polynomials in
distinct $\la \si_y, \si_z\ra$-orbits. We recall the criterion on the $(\Delta_y, \Delta_z)$-exactness of bivariate rational functions
by combining Lemma 3.2 and Theorem 3.3 in~\cite{HouWang2015}.
\begin{lemma}\label{LEM:SSExact}
Let $f\in \bF(y, z)$ be of the form~\eqref{EQ:reduceSS}. Then
$f$ is $(\Delta_y, \Delta_z)$-exact in $\bF(y, z)$ if and only if for all $i,j$ with $1\leq i \leq I$, $1\leq j\leq J_i$,
we have $\si_y^{m_i}(d_i) = \si_z^{n_i}(d_i)$ for some $m_i, n_i\in \bZ$ with $m_i > 0$ and $a_{i, j} = \si_y^{m_i}\si_z^{-n_i}(b_{i, j})-b_{i, j}$
for some $b_{i, j} \in \bF(y)[z]$ with $\deg_z(b_{i, j})<\deg_z(d_i)$.
In particular, if $f$ is $(\Delta_y, \Delta_z)$-exact, so is each $a_{i, j}/d_i^j$.
\end{lemma}

For a rational function $f\in \bF(y, z)$, Abramov's reduction in $z$ and its $q$-analogue in $y$
decompose $f$ into
\begin{equation}\label{EQ:reduceST}
f = \Delta_{q, y}(g)+ \Delta_z(h) + \sum_{i=1}^I\sum_{j=1}^{J_i}\frac{a_{i, j}}{d_i^j},
\end{equation}
where $g, h\in \bF(y, z)$, $a_{i, j}\in \bF(y)[z]$, $d_i\in \bF[y, z]$ satisfy that $\deg_z(a_{i, j})<\deg_z(d_i)$
and $d_i$'s are irreducible polynomials in distinct $\la\tau_{q, y}, \si_z\ra $-orbits.
We recall the criterion on the $(\Delta_{q, y}, \Delta_z)$-exactness in~$\bF(y, z)$ from~\cite[Theorem 3]{Chen2018}.
\begin{lemma}\label{LEM:STExact}
Let $f\in \bF(y, z)$ be of the form~\eqref{EQ:reduceST}.
Then $f$ is $(\Delta_{q, y}, \Delta_z)$-exact in $\bF(y, z)$  if and only if for each $i\in \{1, \ldots, I\}$,
$d_i \in \bF[z]$ and for each $j\in \{1, \ldots,J_i\}$, $a_{i, j} = \Delta_{q, y}(b_{i, j})$ for some $b_{i, j} \in \bF(y)[z]$.
In particular, if $f$ is $(\Delta_{q, y}, \Delta_z)$-exact, so is each $a_{i, j}/d_i^j$.
\end{lemma}

The $q$-analogue of Abramov's reduction decomposes $f\in \bF(y, z)$ into the form
\begin{equation}\label{EQ:qreduceA}
  f = \Delta_{q, z}(g) + c + \frac{a}{b},
\end{equation}
where $g\in \bF(y, z)$, $c\in \bF(y)$ and $a, b\in \bF(y)[z]$ with $\gcd(a, b)=1$, $\deg_z(a)<\deg_z(b)$
and $b$ being $q$-shift-free in $z$ over $\bF(y)$, that is gcd$(b,\tau_{q,z}^kb)=1$ for any $k\in\bZ\setminus\{0\}$. Moreover, $f = \Delta_{q, z}(u)$ for some $u\in \bF(y, z)$
if and only if $c=0$ and $a=0$.

Applying the reduction formula~\eqref{eq:red} with $\theta= \tau_{q,y}$, we can further decompose $f$ as
\begin{equation}\label{EQ:reduceTT}
  f = \Delta_{q,y}(u) + \Delta_{q,z}(v) +c+ \sum_{i=1}^I\sum_{j=1}^{J_i} \frac{a_{i, j}}{d_i^j},
\end{equation}
where $u, v\in \bF(y, z)$, $c\in\bF(y)$, $a_{i, j}\in \bF(y)[z]$, and $d_i\in \bF[y, z]$ are such that
$\deg_z(a_{i, j})<\deg_z(d_i)$ and the $d_i$'s are irreducible  polynomials in
distinct $\la \tau_{q,y}, \tau_{q,z}\ra$-orbits.
Then the $(\Delta_{q,y}, \Delta_{q,z})$-exactness criterion of $f$ can be given
by combining Lemma 3.6 and Theorem 3.8 in~\cite{Wang2020}, which is a $q$-analogue of Lemma~\ref{LEM:SSExact}.
\begin{lemma}\label{LEM:TTExact}
Let $f\in \bF(y, z)$ be of the form~\eqref{EQ:reduceTT}.
Then $f$ is $(\Delta_{q,y}, \Delta_{q,z})$-exact in $\bF(y, z)$ if and only if
$c=\Delta_{q,y}(h)$ for some $h\in\bF(y)$
and for all $i,j$ with $1\leq i \leq I$, $1\leq j\leq J_i$, we have $\si_y^{m_i}(d_i) = q^{s_i}\si_z^{n_i}(d_i)$ for some~$m_i, n_i, s_i\in \bZ \text { with } m_i > 0$
and for the smallest positive integer $m_i$, $a_{i, j} = q^{-js_i}\tau_{q,y}^{m_i}\tau_{q,z}^{-n_i}(b_{i, j})-b_{i, j}$
for some $b_{i, j} \in \bF(y)[z]$ with $\deg_z(b_{i, j})<\deg_z(d_i)$.
In particular, if $f$ is $(\Delta_{q,y}, \Delta_{q,z})$-exact, so is each $a_{i, j}/d_i^j$.
\end{lemma}

\subsection{The mixed cases}\label{SSUBSECT:mixed}

For a rational function $f\in \bF(y, z)$, applying the Ostrogradsky--Hermite reduction in~$z$ and the reduction
formula~\eqref{eq:red} with $\theta = \theta_y\in \{\si_y, \tau_{q, y}\}$
to $f$ yields
\begin{equation}\label{EQ:reduceSD}
  f = \Theta_y(u) + D_z(v) + \sum_{i=1}^I \frac{a_i}{d_i},
\end{equation}
where $u, v \in \bF(y, z), a_{i}\in \bF(y)[z], d_i\in \bF[y, z]$ with $\deg_z(a_{i})< \deg_z(d_i)$
and the $d_i$'s are irreducible polynomials in distinct $\la\theta_y\ra $-orbits.
We recall the criterion on the $(\Theta_y, D_z)$-exactness in~$\bF(y, z)$ from~\cite[Theorem 2]{Chen2018}.
\begin{lemma}\label{LEM:SDExact}
Let $\theta_y \in \{\si_y, \tau_{q, y}\}$ and $f\in \bF(y, z)$ be of the form~\eqref{EQ:reduceSD}.
Then $f$ is $(\Theta_y, D_z)$-exact in $\bF(y, z)$  if and only if for each $i\in \{1, \ldots, I\}$,
$d_i \in \bF[z]$ and $a_i = \Theta_y(b_i)$ for some $b_i \in \bF(y)[z]$.
In particular, if $f$ is $(\Theta_y, D_z)$-exact, so is each $a_i/d_i$.
\end{lemma}



\section{Existence Criteria}\label{SECT:criteria}

We will reduce the existence problem of telescopers in the trivariate case
to that in the bivariate case and two related problems. To this end,
we first recall the existence criteria
on telescopers for bivariate rational functions
from~\cite{AbramovLe2002, Le2001, Abramov2003, ChenSinger2012, Chen2015}.

\begin{theorem}\label{THM:bicriteria}
Let $f(x, y)$ be a rational function in $\bK(x, y)$. Then
\begin{enumerate}[(i)]
\item Differential case (see~\cite[Theorem 4.5]{ChenSinger2012}): $f$ always has a telescoper of type $(D_x, D_y)$;
\item Shift case (see~\cite[Theorem 1]{AbramovLe2002} or~\cite[Theorem 4.11]{ChenSinger2012}): $f$ has a telescoper of type $(S_x, \Delta_y)$ if and only if $f$
is of the form $f = \Delta_y(g) + r$ for some $g, r\in \bK(x, y)$ and $r$ is proper in $\bK(x, y)$.
\item $q$-Shift case (see~\cite[Theorem 1]{Le2001} or~\cite[Theorem 4.15]{ChenSinger2012}):
$f$ has a telescoper of type $(T_{q, x}, \Delta_{q, y})$ if and only if $f$
is of the form $f = \Delta_{q, y}(g) + r$ for some $g, r\in \bK(x, y)$ and $r$ is $q$-proper in $\bK(x, y)$.
\item Mixed cases (see~\cite[Theorems 4.6, 4.7, 4.9, 4.12, 4.13, 4.14]{ChenSinger2012}): $f$ has a telescoper of type $(\partial_x, \Theta_y) \in \{(S_x, D_y), (T_{q, x}, D_y),
(D_x, \Delta_y), (T_{q, x}, \Delta_y), (D_x, \Delta_{q, y}), (S_x, \Delta_{q, y})\}$ if and only if $f$
is of the form $f = \Theta_y(g) + r$ for some $g, r\in \bK(x, y)$ and the denominator of $r$ is split
with respect to the partition $(\{x\}, \{y\})$.
\end{enumerate}
\end{theorem}
\begin{example}
Let $f = 1/(x+y)$. Then $f$ has a telescoper of type $(D_x, D_y)$, $(S_x, \Delta_y)$ and $(T_{q, x}, \Delta_{q, y})$, but has
no telescoper in the mixed cases since $x+y$ is not split.
\end{example}

\begin{problem}[Shift Equivalence Testing Problem]\label{PROB:shiftequiv}
Let $\bF$ be any computable field of characteristic zero.
Given $p \in \bF[x_1, ..., x_n]$, decide whether there exist $m_1, \ldots, m_n$ $\in \bZ$ with
$m_1 >0$ such that $p(x_1 + m_1, \ldots, x_n+m_n)= p(x_1, \ldots, x_n)$.
\end{problem}
This problem is solved by Grigoriev in~\cite{Grigoriev1996,Grigoriev1997} and more recently by Kauers and Schneider in~\cite{Kauers2010}
and Dvir et al. in~\cite{Dvir2014}.

\begin{problem}[$q$-Shift Equivalence Testing Problem]\label{PROB:qshiftequiv}
Let $p \in \bF[x_1, ..., x_n]$, decide if there exist $m, m_1, \ldots, m_n$ $\in \bZ$ with
$m_1 >0$ such that $p(q^{m_1}x_1, \ldots, q^{m_n}x_n)= q^{m}p(x_1, \ldots, x_n)$.
\end{problem}
This problem is much easier than the shift case, and an algorithm for testing $q$-shift equivalence has been given in~\cite{Wang2020}.

\begin{problem}[Separation Problem]\label{PROB:sep}
Given an algebraic function $\alpha\in \overline{\bK(x, y)}$, decide whether
there exists a nonzero operator $L\in \bK(x)\la D_x \ra$ such that $L(\alpha)=0$.
If such an operator exists, we say that $\alpha$ is \emph{separable} in $x$ and $y$.
\end{problem}

As a special case of~\cite[Proposition 10]{Chen2014}, a rational function in $\bK(x,y)$ is separable
if and only if it is of the form $a/(bc)$ with $a\in \bK[x, y], b\in \bK[x]$ and $c\in \bK[y]$.
This motivates the nomenclature of Problem~\ref{PROB:sep}.
We will study the separation problem in the forthcoming paper~\cite{Chen2019}, in which an algorithm is presented for constructing such a differential annihilator $L\in \bK(x)\la D_x \ra$ if it exists.

\subsection{Existence problems of first class }\label{SUBSECT:class1}

In the pure differential setting, telescopers always exist for general $D$-finite functions
over $\bK(\vv)$, which was proved by Zeilberger in 1990 using the elimination property of holonomic D-modules~\cite{Zeilberger1990}.
For the sake of completeness, we will give a more direct proof for rational functions in $\bK(\vv)$.
We first adapt Wegschaider's \lq\lq non-commutative trick\rq\rq~in~\cite[Theorem~3.2]{Wegschaider1997} to the differential case.
\begin{lemma}\label{LM:telexist}
Let~$f\in \bK(x, y_1, \ldots, y_n)$ and~$A \in \bK[x]\langle D_x, D_{y_1}, \ldots, D_{y_n}\rangle$ be a nonzero operator such that~$A(f)=0$.
Then there exists a nonzero operator~$L\in \bK[x]\langle D_x \rangle$
such that~$L(f)= D_{y_1}(g_1) + \cdots + D_{y_n}(g_n)$ for some~$g_1, \ldots, g_n\in \bK(x, y_1, \ldots, y_n)$.
\end{lemma}
\begin{proof} We will follow the same argument in the proof of~\cite[Theorem~3.2]{Wegschaider1997}.
We claim that for every $\ell \in \{1, \ldots, n+1\}$, there exist $Q_{j,\ell}\in \bK[\vv]\la D_x, D_{y_j}, \ldots, D_{y_n}\ra$
for each $j\in \{1, \ldots, \ell-1\}$ and a nonzero $R_\ell \in \bK[x]\la D_x, D_{y_\ell}, \ldots, D_{y_n}\ra$ such that $f$
is annihilated by the operator
\begin{equation}\label{eq:operator}
P_\ell = \sum_{j=1}^{\ell-1} D_{y_j}Q_{j,\ell} + R_\ell.
\end{equation}
The lemma follows from this claim since $R_{n+1}$ is the desired operator $L\in \bK[x]\langle D_x \rangle$ with
$g_j := -Q_{j,\ell}(f) \in \bK(\vv)$ for $j\in \{1, \ldots, n\}$.

We prove the claim inductively: for $\ell=1$ take $P_1 = R_1 = A$. Assume that for some
$\ell \in \{1, \ldots, n\}$ we have a nonzero operator $P_\ell$ of the form~\eqref{eq:operator}
that annihilates $f$. We show that by division of $R_\ell$ by $D_{\ell}$ we can construct the operator $P_{\ell+1}$.

Since~$D_{y_\ell}$ commutes with~$x$ and~$D_x, D_{y_{\ell+1}}, \ldots, D_{y_n}$,
we can write~$R_{\ell} = D_{y_\ell}^m(R_{\ell+1}+ D_{y_\ell} M)$,
where~$m\in \bN$, ~$M\in \bK[x]\langle D_x, D_{y_\ell}, \ldots, D_{y_n}\rangle$, and~$R_{\ell+1}$
is a nonzero operator in~$\bK[x]\langle D_x, D_{y_{\ell+1}}, \ldots, D_{y_n} \rangle$. For any~$w \in
\bK[y_{\ell}]$ of degree at most~$m$, we have
\begin{equation} \label{EQ:ddrem}
w D_{y_\ell}^m = D_{y_\ell} \tilde{Q}_{\ell} + r
\end{equation}
for some~$r \in \bK$ and~$\tilde{Q}_\ell \in \bK[y_\ell]\langle D_{y_\ell} \rangle$.  In particular, ~$r = (-1)^m m!\neq 0$ if we take~$w = y_{\ell}^m$.
Using the fact~$r D_{y_i} = D_{y_i} r$ for all $i\in \{1, \ldots, n\}$ and~\eqref{EQ:ddrem}, we find
\begin{align*}
   \frac{y_{\ell}^m}{(-1)^m m!} P_{\ell} & = \sum_{j=1}^{\ell-1} D_{y_j}
\left(\frac{y_{\ell}^m}{(-1)^m m!} Q_{j,\ell}\right) + \frac{y_{\ell}^m}{(-1)^m m!} D_{y_\ell}^m (R_{\ell+1}+ D_{y_\ell} M)\\
   & = \sum_{j=1}^{\ell-1} D_{y_j}
\left(\frac{y_{\ell}^m}{(-1)^m m!} Q_{j,\ell}\right) + \left(D_{y_\ell}\tilde{Q}_\ell + 1\right)(R_{\ell+1}+ D_{y_\ell} M)\\
& = \sum_{j=1}^{\ell} D_{y_j} \tilde{Q}_{j,\ell}  + R_{\ell+1} \triangleq P_{\ell+1} \quad \text{with~$\tilde Q_{j,\ell}\in \bK[\vv]\la D_x, D_{y_j}, \ldots, D_{y_n}\ra$.}
\end{align*}
Since~$P_\ell(f)=0$, we have $P_{\ell+1}(f)=0$. So $P_{\ell + 1}$ is the desired operator.
\end{proof}

\begin{theorem}\label{THM:diff}
For any rational function~$f\in \bK(\vv)$, there exists a nonzero~$L\in \bK[x]\langle D_x \rangle$
such that~$L(f)= D_{y_1}(g_1) + \cdots + D_{y_n}(g_n)$ for some~$g_1, \ldots, g_n\in \bK(\vv)$.
\end{theorem}
\begin{proof} It suffices to show that there exists a nonzero $A \in \bK[x]\langle D_x, D_{y_1}, \ldots, D_{y_n}\rangle$
such that $A(f)=0$ by Lemma~\ref{LM:telexist}.
Write~$f=P/Q$ with~$P, Q\in \bK[\vv]$ and~$\gcd(P, Q)=1$.
Denote~$d_x = \max\{\deg_x(P), \deg_x(Q)\}$ and~$d_{y_i} = \max\{\deg_{y_i}(P), \deg_{y_i}(Q)\}$ for $i\in \{1, \ldots, n\}$.
Let $\bW_N$ be the $\bK$-vector space generated by the set
\[\{\,x^iD_x^{j_0}D_{y_1}^{j_1}\cdots D_{y_n}^{j_n} \mid 0\leq i+j_0+\cdots + j_n \leq N\,\}\]
over~$\bK$.
By an easy combinatorial counting, the
dimension of~$\bW_N$ is~$\binom{N+n+2}{n+2} = \bigO(N^{n+2})$ over~$\bK$.
Furthermore, for any $(i, j_0, \ldots, j_n) \in \bN^{n+2}$, a direct calculation yields
\begin{equation}\label{eq:nonminimal}
x^iD_x^{j_0}D_{y_1}^{j_1}\cdots D_{y_n}^{j_n}(f)=\frac{P_{i,j_0, \ldots, j_n}}{Q^{i+j_0+\cdots+j_n+1}},
\end{equation}
where~$P_{i, j_0, \ldots, j_n}\in \bK[\vv]$ with $\deg_x(P_{i,j_0, \ldots, j_n})\leq
(i+j_0+\cdots+j_n+1)d_x+i$ and \[\deg_{y_i}(P_{i,j_0, \ldots, j_n})\leq (i+j_0+\cdots+j_n+1)d_{y_i} \quad \text{for $i\in \{1, \ldots, n\}$}.\]
So the set~$\bW_N(f)$ is included in the $\bK$-vector space $\bV_N$ spanned by the set
\[\left\{ \frac{x^{k_0} y_1^{k_1}\cdots y_n^{k_n}}{Q^{N+1}} \middle\vert 0\leq k_0\leq (N+1)d_x+N, \, 0\leq k_i\leq (N+1)d_{y_i}\, \text{for $i=1, \ldots, n$}  \right\},\]
whence the dimension of~$\bV_N$ is~$(N+1)(d_x + 1)\prod_{i=1}^n ((N+1)d_{y_i} + 1) = \bigO(N^{n+1})$ over~$\bK$. Define
linear map~$\psi: \bW_N\rightarrow \bV_N$ by~$\psi(L)=L(f)$ for any~$L\in \bW_N$.
For sufficiently large $N$, we have
\[\binom{N+n+2}{n+2} > (N+1)(d_x + 1)\prod_{i=1}^n ((N+1)d_{y_i} + 1),\]
which implies that the kernel of~$\psi$ is nontrivial.
Therefore, there exists a nonzero operator~$A\in \bW_N \subseteq \bK[x]\langle D_x, D_{y_1}, \ldots, D_{y_n}\rangle$
such that $A(f)=0$.
\end{proof}

\begin{remark}
In the continuous setting, the existence of telescopers for rational functions
implies that for algebraic functions by~\cite[Lemma 4]{Chen2012}. Efficient algorithms for computing
telescopers have been given in~\cite{BCCL2010, Chen2012, BLS2013, Lairez2016}.
\end{remark}


\subsection{Existence problems of second class }\label{SUBSECT:class2}
{We now solve the second class of existence problems
where telescopers are linear differential operators in $\bK(x)\la D_x \ra$ and $(\Theta_y,\Theta_z)\in \{(\Delta_y,\Delta_z),
(\Delta_{q,y},\Delta_{z}),(\Delta_{q,y},\Delta_{q,z})\}$.
\begin{problem}\label{Problem:type2}
Given $f\in\bK(x,y,z)$, determine if there exists a nonzero operator $L\in\bK(x)\la D_x \ra$ such that
$L(f)=\Theta_y(g)+\Theta_z(h)$ for some $g,h\in\bK(x,y,z)$.
\end{problem}
For $v\in\{y,z\}$, let $\theta_v=\sigma_v$ if $\Theta_v=\Delta_v$ or $\theta_v=\tau_{q,v}$ if $\Theta_v=\Delta_{q,v}$.
By partial fraction decomposition w.r.t $z$ and the transformation~\eqref{eq:red} with $\theta=\theta_y$ and subsequently with $\theta=\theta_z$, any rational function $f\in\bK(x,y,z)$ can be decomposed into
\begin{equation}\label{eq:reduced form1}
f=\Theta_y(u)+\Theta_z(v)+\mu+\sum_{i=1}^{I}\sum_{j=1}^{J_i}\frac{a_{i,j}}{d_i^j},
\end{equation}
where $u,v\in\bK(x,y,z),\mu\in\bK(x,y),a_{i,j}\in\bK(x,y)[z],d_i\in\bK[x,y,z]$ with $\deg_z(a_{i,j})<\deg_z(d_i)$, $d_i$'s are irreducible polynomials in distinct $\la\theta_y,\theta_z\ra$-orbits and none of nonzero $a_{i,j}/d_i^j$ is $(\Theta_y,\Theta_z)$-exact.

The following theorem shows that Problem \ref{Problem:type2} can be reduced to the same problem but for simple fractions and bivariate rational functions.
\begin{theorem}\label{T2Reduce}
Let $f\in\bK(x,y,z)$ be of the form \eqref{eq:reduced form1}.
Then $f$ has a telescoper of type $(D_x,\Theta_y,\Theta_z)$ if and only if $\mu$ and the fraction $a_{i,j}/d_i^j$ has a telescoper of the same type for all $i,j$ with $1\leq i\leq I$ and $1\leq j\leq J_i$.
\end{theorem}
\proof The sufficiency follows from Lemma \ref{LM:closure}.
For the necessity, when $f$ has a telescoper of type $(D_x,\Theta_y,\Theta_z)$, since $D_x$ does not change the $\la\theta_y,\theta_z\ra$-equivalence of the denominators, one can deduce that $\mu$ and $r=\sum_{i=1}^{I}\sum_{j=1}^{J_i}\frac{a_{i,j}}{d_i^j}$ both have a telescoper of the same type.

Next we will show each fraction $a_{i,j}/d_i^j$ has a telescoper of the same type when $r$ has a telescoper.
To this end, we first show that $D_x(d_i)=0$, that is $d_i\in\bK[y,z]$ for all $1\leq i\leq I$.
Over the field $\overline{\bK(x,y)}$, we can decompose $r$ as
\[
  r=\Theta_y(u^{\star})+\Theta_z(v^{\star})+r^{\star}\text{ with }r^{\star}=\sum_{i=1}^{I'}\sum_{j=1}^{J_i'}
  \frac{\alpha_{i,j}}{(z-\beta_i)^j},
\]
where $u^{\star},v^{\star}\in\overline{\bK(x,y)}(z)$, $\alpha_{i,j},\beta_i\in\overline{\bK(x,y)}$ with $\alpha_{i,J_i'}\neq 0$, $z-\beta_i$ and $z-\beta_{i'}$ are not $\la\theta_y,\theta_z\ra$-equivalent for all $i,i'$ with $1\leq i\neq i'\leq I'$.
It suffices to show $D_x(\beta_i)=0$ for all $i$ with $1\leq i\leq I'$.
We will prove this claim by contradiction.
Suppose that $D_x(\beta_k)\neq 0$ for some $1\leq k\leq I'$ and that $L=\sum_{\ell=0}^{\rho}e_{\ell}D_x^{\ell}\in\bK(x)\la D_x\ra$ with $e_{\rho}e_0\neq 0$ is a telescoper for $r^{\star}$.
Then
\[
  L(r^{\star})=\sum_{i=1}^{I'}\left(
  \frac{J_i'^{\overline{\rho}}e_{\rho}\alpha_{i,J_i'}D_x(\beta_i)^{\rho}}
       {(z-\beta_i)^{J_i'+\rho}}
  +\sum_{j=1}^{J_i'+\rho-1}\frac{\tilde{\alpha}_{i,j}}{(z-\beta_i)^j}\right),
\]
where $J_i'^{\overline{\rho}}=J_i'(J_i'+1)\cdots(J_i'+\rho-1)$ and $\tilde{\alpha}_{i,j}\in\overline{\bK(x,y)}$.
As $L(r^{\star})$ is $(\Theta_y,\Theta_z)$-exact and $D_x(\beta_k)\neq 0$,
we have
\begin{equation}\label{eq:1condition}
\theta_y^{m_k}(z-\beta_k)=q^{s_k}\theta_z^{n_k}(z-\beta_k)
\end{equation}
for some $m_k,n_k,s_k\in\bZ$ with $m_k>0$ and
\begin{equation}\label{eq:2condition}
  J_k'^{\overline{\rho}}e_{\rho}\alpha_{k,J_k'}D_x(\beta_k)^{\rho}
  =q^{-(J_k'+\rho)s_k}\theta_y^{m_k}(\gamma_k)-\gamma_k
\end{equation}
for some $\gamma_k\in\overline{\bK(x,y)}$.
From the Equation \eqref{eq:1condition}, we know $\theta_y^{m_k}D_x(\beta_k)=q^{s_k}D_x(\beta_k)$.
Dividing Identity \eqref{eq:2condition} by $J_k'^{\overline{\rho}}
            e_{\rho}D_x(\beta_k)^{\rho}$ gives
\[
  \alpha_{k,J_k'}=q^{-J_k's_k}\theta_y^{m_k}\left(\frac{\gamma_k}
  {J_k'^{\overline{\rho}}
            e_{\rho}D_x(\beta_k)^{\rho}}\right)
            -\frac{\gamma_k}{J_k'^{\overline{\rho}}
            e_{\rho}D_x(\beta_k)^{\rho}}.
\]
Thus $\frac{\alpha_{k,J_k'}}{(z-\beta_k)^{J_k'}}$ is $(\Theta_y,\Theta_z)$-exact in $\overline{\bK(x,y)}(z)$, and hence can be moved into $u^{\star}$ and $v^{\star}$.
Then by similar discussions as above, one can see $\frac{\alpha_{k,j}}{(z-\beta_k)^{j}}$ is $(\Theta_y,\Theta_z)$-exact for all $j$ with $1\leq j\leq J_i$.
Notice that $\beta_k$ is a root of $d_k$ for some $1\leq k\leq I$ and that
$D_x(\beta_k)\neq 0$ leads to $D_x(\beta)\neq 0$ for any conjugate root $\beta$ of $d_k$.
Then all fractions of the form $\frac{\alpha}{(z-\beta)^j}$ in $r^{\star}$ are
also $(\Theta_y,\Theta_z)$-exact.
Collecting all these fractions together, we get $\frac{a_{k,j}}{d_k^j}$ is $(\Theta_y,\Theta_z)$-exact in $\overline{\bK(x,y)}(z)$ and hence in $\bK(x,y,z)$ by Lemma \ref{LEM:trace}, which contradicts the assumption that none of nonzero $\frac{a_{i,j}}{d_i^j}$ in $r$ is not exact.
At this stage we have proved $d_i\in\bK[y,z]$.
Since $L$ is also a telescoper for $r$. Then
\[
  L(r)=\sum_{i=1}^{I}\sum_{j=1}^{J_i}\frac{L(a_{i,j})}{d_i^j}=\Theta_y(g)+\Theta_z(h)
\]
for some $g,h\in\bK(x,y,z)$.
Notice that $d_i$'s are in distinct $\la\theta_y,\theta_z\ra$-orbits,
\[
L\left(\frac{a_{i,j}}{d_i^j}\right)=\frac{L(a_{i,j})}{d_i^j} =\Theta_y(g_{i,j})+\Theta_z(h_{i,j})
\]
for some $g_{i,j},h_{i,j}\in\bK(x,y,z)$.
So $L$ is a telescoper for all $a_{i,j}/d_i^{j}$ with $1\leq i\leq I$ and $1\leq j \leq J_i$.
\qed

Notice that for $\mu\in\bK(x,y)$, having telescopers of type $(D_x,\Theta_y,\Theta_z)$ or $(D_x,\Theta_y)$ are equivalent.
As the existence problem of telescopers for bivariate rational functions has been settled by Theorem \ref{THM:bicriteria},
we only need to decide when $f=\frac{a}{d^j}$,
where $a\in\bK(x,y)[z], d\in\bK[x,y,z]$ with $\deg_z(a)<\deg_z(d)$ and $d$ being irreducible, has telescopers of type $(D_x,\Theta_y,\Theta_z)$.
Same argument as in the proof of Theorem \ref{T2Reduce} implies that $f$ has a telescoper of type $(D_x,\Theta_y,\Theta_z)$ only when $d$ is free of $x$. Assume $d\in\bK[y,z]$ and $L\in\bK(x)\la D_x\ra$ is a telescoper of $f$.
Then $L(f)=\frac{L(a)}{d^j}$ is $(\Theta_y,\Theta_z)$-exact.
We will proceed by checking whether the two conditions for the exactness in
Lemma \ref{LEM:SSExact}, \ref{LEM:TTExact} and \ref{LEM:STExact} are satisfied.

If $\theta_y^{m}(d)\neq q^{t}\theta_z^{n}(d) \text{ whenever } m,n,t\in\mathbb{Z} \text{ and } m>0$, then we have $L(a)=0$ which can be reduced to solving the separation problem of bivariate rational functions and settled via GCD computations.

If $\theta_y^{m}(d)=q^{t}\theta_z^{n}(d) \text{ for some } m,n,t\in\mathbb{Z} \text{ with } m$ being the smallest positive integer, then
$L(a)$ satisfies an equation.
Next we will show how to solve the equation for different $(\theta_y,\theta_z)$ separately.
\begin{itemize}
\item[{\rm (1)}]
When $(\theta_y, \theta_z)=(\sigma_y, \si_z)$.
Lemma \ref{LEM:SSExact} shows that $L(x, D_x)(a)=\sigma_y^{m}\sigma_z^{-n}(b)-b$ for some $b\in\bK(x,y)[z]$ with $\deg_z(b)<\deg_z(d)$.
Taking $\bar{y}=y/m$ and $\bar{z}=ny+mz$ shows
$L(x, D_x)(a)=\sigma_y^{m}\sigma_z^{-n}(b)-b$ is equivalent to the existence problem of telescopers of type $(D_x,\Delta_y)$ for bivariate rational functions, which has been solved by Theorem \ref{THM:bicriteria}.
\item[{\rm (2)}]
When $(\theta_y, \theta_z)=(\tau_{q,y}, \si_z)$, Lemma \ref{LEM:STExact} leads to $L(a)=\Delta_{q,y}(b)$, which is the existence problem of telescopers  of type $(D_x,\Delta_{q,y})$ solved by Theorem \ref{THM:bicriteria}.
\item[{\rm (3)}]
When $(\theta_y, \theta_z)=(\tau_{q,y}, \tau_{q, z})$, by Lemma \ref{LEM:TTExact} we know
$L(x, D_x)(a)=q^{-j t}\tau_{q,y}^{m}\tau_{q,z}^{-n}(b)-b$ for some $b\in\bK(x,y)[z]$ with $\deg_z(b)<\deg_z(d)$.
Define an $\bF$-homomorphism $\varphi$ of $\bF(y,z)$ by $y\mapsto y^{m},z\mapsto y^{-n}z$. Then the $q$-difference equation can also be simplified.
\begin{proposition}
Given a rational function $f\in\bF(y,z)$ and integers $m,n,s\in\mathbb{N}$ with $m>0$.
Then $f=q^{s}\tau_{q,y}^m\tau_{q,z}^{-n}(g)-g$ for some $g\in\bF(y,z)$ if and only if $\varphi(f)=q^s\tau_{q,y}(h)-h$ for some $h\in\bF(y,z)$.
\end{proposition}
\proof Let $\tau=\tau_{q,y}^m\tau_{q,z}^{-n}$.
The necessity follows from the fact that $\varphi \circ \tau=\tau_{q,y}\circ \varphi$.
For the sufficiency, define $\psi:\bF(y,z)\rightarrow \overline{\bF(y,z)}$ by $y\mapsto y^{1/m}, z\mapsto y^{n/m}z$, where $\overline{\bF(y,z)}$ is the algebraic closure of $\bF(y,z)$.
It is easy to see $\psi\circ \varphi=Id_{\bF(y,z)}$ and $\psi \circ\tau_{q,y}=\tau\circ\psi$, where $\tau_{q,y}$ is extended to $\overline{\bF(y,z)}$.
Thus $\varphi(f)=q^{s}\tau_{q,y}(h)-h$ implies $f=q^{s}\tau_{q,y}^m\tau_{q,z}^{-n}(\tilde{g})-\tilde{g}$ with $\tilde{g}=\psi(h)\in\overline{\bF(y,z)}$.
By similar trace arguments used in Lemma \eqref{LEM:trace}, one can see
$f=q^{s}\tau_{q,y}^m\tau_{q,z}^{-n}(\tilde{g})-\tilde{g}$ if and only if $f=q^{s}\tau_{q,y}^m\tau_{q,z}^{-n}(g)-g$ for some $g\in\bF(y,z)$.
\qed

At this stage, by letting $\bar{y}=y^{1/m}$ and $\bar{z}=y^{n/m}z$, we only need to decide whether $L(\bar{a})=q^{-jt}\tau_{q,y}(\bar{b})-\bar{b}$ for some $\bar{b}\in\bK(x,y,z)$, which can be determined by a similar discussion process as the existence problem of telescopers of type $(D_x,\Delta_{q,y})$.
\end{itemize}
}

\subsection{Existence problems of third class }\label{SUBSECT:class3}

We now consider the third class of the existence problems of telescopers for
rational functions in three variables.

\begin{problem}\label{PROB:SDD}
Given $f\in \bK(x, y, z)$, decide whether there exists a nonzero operator $L$ in $\bK(x)\langle \partial_x \rangle$ with $\partial_x\in \{S_x, T_{q, x} \}$
such that~$L(f) = D_y(g) + D_z(h)$
for some $g, h\in \bK(x, y, z)$.
\end{problem}

Let $f\in \bF(y, z)$ be of the form~\eqref{EQ:reduceOH} with $\bF = \bK(x)$.
If $f$ is $(D_y, D_z)$-exact in $\bK(x, y, z)$, then $1$ is a telescoper for $f$.
From now on, we assume that $f$ is not $(D_y, D_z)$-exact. Let $(\partial_x,\theta_x) \in \{(S_x,\si_x), (T_{q,x},\tau_{q, x})\}$.
By dividing the roots of $b$ in $\overline{\bK(x,y)}$ into different $\la \theta_x\ra$-orbits, we can write $f$ as $f = D_z(u) + r$ with $u\in \bF(y, z)$
and
\begin{equation}\label{EQ:SDDred2}
 r = \sum_{i=1}^I \sum_{j=0}^{J_i} \frac{\alpha_{i, j}}{z-\theta_x^j(\beta_i)},
\end{equation}
where $\alpha_{i, j}, \beta_i \in \overline{\bK(x, y)}$ and the $\beta_i$'s are in distinct $\la \theta_x\ra$-orbits.
Note that $f$ has a telescoper of type $(\partial_x, D_y, D_z)$ if and only if $r$ has a telescoper of the same type.

\begin{lemma} \label{LEM:SDDlocal}
Let $r = \sum_{j=0}^J {\alpha_j}/{(z-\theta_x^j(\beta))}$ with $\alpha_j, \beta\in \overline{\bK(x, y)}$ and
$\theta_x^m(\beta)\neq \beta$ for any $m\in \bZ\setminus \{0\}$.
Then $r$ is $(D_y, D_z)$-exact if it has a telescoper of type $(\partial_x, D_y, D_z)$.
\end{lemma}
\begin{proof}
Assume that $L = \sum_{\ell=0}^{\rho} e_{\ell} \partial_x^{\ell} \in \bK(x)\la \pa_x\ra$ with $e_0\neq 0$ is a telescoper for $r$
of type $(\partial_x, D_y, D_z)$. Then
\[L(r) = \sum_{j=0}^{J + \rho} \frac{\tilde \alpha_{j}}{z-\theta_x^j(\beta)} = D_y(u) + D_z(v),\]
where $u, v \in \overline{\bK(x, y)}(z)$ and $\tilde{\alpha}_{j} = \sum_{k=0}^j e_{k}\theta_x^k(\alpha_{j-k})$ with $e_k = 0$ for $k>\rho$
and $\alpha_{j}=0$ for $j > J$.
Since $\theta_x^m(\beta)\neq \beta$ whenever $m\in \bZ\setminus \{0\}$, for each $1\leq j\leq J+\rho$ we have
$\tilde \alpha_j = D_y(\tilde \gamma_j)$ for some $\tilde \gamma_j \in \overline{\bK(x, y)}$ by Lemma~\ref{LEM:DDExact}.
We now prove inductively that for each $j$ with $0\leq j \leq J$, $\alpha_j = D_y(\gamma_j)$ for some  $\gamma_j \in \overline{\bK(x, y)}$.
Since $\tilde \alpha_0 = e_0 \alpha_0$ and $e_0\in \bK(x)\setminus \{0\}$, we have $\alpha_0 = D_y(\gamma_0)$ with $\gamma_0 = \tilde \gamma_0/e_0$.
Suppose that we have shown that $\alpha_j = D_y(\gamma_j)$ for $j=0, \ldots, k-1$ with $k\leq J$.
Note that $\tilde \alpha_k = e_0 \alpha_k + e_1 \theta_x(\alpha_{k-1}) + \cdots + e_k\theta_x^k(\alpha_0) = D_y(\tilde \gamma_k)$.
Then $\alpha_k = D_y(\gamma_k)$ with $\gamma_k = \frac{1}{e_0}(\tilde \gamma_k - \sum_{j=1}^k e_j \theta_x^j(\gamma_{k-j}))$.
So $r$ is $(D_y, D_z)$-exact by Lemma~\ref{LEM:DDExact}.
\end{proof}

\begin{theorem}\label{THM:SDDcriteria}
Let $r\in \bK(x, y, z)$ be of the form~\eqref{EQ:SDDred2}.
Then $r$ has a telescoper of type $(\partial_x, D_y, D_z)$ if and only if for each $i$ with $1\leq i \leq I$, either $\alpha_{i, j}/(z-\theta_x^j(\beta_i))$
is $(D_y, D_z)$-exact or $\beta_i\in \overline{\bK(y)}$ and there exists a nonzero $L_{i, j}\in \bK(x)\la \partial_x \ra$ such that
$L_{i, j}(\alpha_{i, j}) = D_y(\gamma_{i, j})$ for some $\gamma_{i, j} \in \bK(x, y)(\beta_i)$.
\end{theorem}
\begin{proof}
The sufficiency follows from Lemma~\ref{LM:closure} since each fraction $\alpha_{i, j}/(z-\theta_x^j(\beta_i))$
is either $(D_y, D_z)$-exact or has a telescoper of type $(\partial_x, D_y, D_z)$. To show the necessity, we assume that
$L = \sum_{\ell=0}^{\rho} e_{\ell} \partial_x^{\ell} \in \bK(x)\la \partial_x\ra$ with $e_0\neq 0$ is a telescoper
for $r$ of type $(\partial_x, D_y, D_z)$.
Then we have
\[L(r) =   \sum_{i=1}^I \sum_{j=0}^{J_i + \rho} \frac{\tilde \alpha_{i, j}}{z-\theta_x^j(\beta_i)} = D_y(u) + D_z(v), \]
where $u, v \in \bK(x, y, z)$ and $\tilde{\alpha}_{i, j} = \sum_{k=0}^j e_{k}\theta_x^k(\alpha_{i, j-k})$ with $e_k = 0$ for $k>\rho$
and $\alpha_{i, j}=0$ for $j > J_i$. By Lemma~\ref{LEM:DDExact}, we have $r_i = \sum_{j=0}^{J_i + \rho} \frac{\tilde \alpha_{i, j}}{z-\theta_x^j(\beta_i)}$
is $(D_y, D_z)$-exact for each $i$ with $1\leq i \leq I$ since the $\beta_i$'s are in distinct $\la \theta_x\ra$-orbits.
If there exists a nonzero $m_i \in \bN$ such that $\theta_x^{m_i}(\beta_i) = \beta_i$, then $\beta_i \in \overline{\bK(y)}$ by~\cite[Lemma 3.4~(i)]{ChenSinger2012}. So
$J_i = 0$  and $L(\alpha_{i, 0}/(z-\beta_i)) = L(\alpha_{i, 0})/(z-\beta_i)$ is $(D_y, D_z)$-exact, which implies that $L(\alpha_{i, 0})= D_y(\gamma_{i, 0})$
for some $\gamma_{i, 0} \in \overline{\bK(x, y)}$. Since $\alpha_{i, 0} \in \bK(x, y)(\beta_i)$, we can choose $\gamma_{i, 0}\in \bK(x, y)(\beta_i)$ by the trace argument.
If there is no nonzero $m_i\in \bN$ such that $\theta_x^{m_i}(\beta_i) = \beta_i$, then the theorem follows from Lemma~\ref{LEM:SDDlocal}.
\end{proof}

Problem~\ref{PROB:SDD} now has been reduced to the exactness testing problem and the following existence problem.

\begin{problem}\label{PROB:SDDsimple}
Given $\alpha\in \bK(x, y)(\beta)$ with $\beta$ algebraic over $\bK(y)$, decide whether $\alpha$ has a telescoper
of type $(\partial_x, D_y)$ with $\partial_x \in \{S_x, T_{q, x}\}$, i.e.,
there exists a nonzero  $L\in \bK(x)\langle \partial_x \rangle$
such that $ L(\alpha) = D_y(\gamma)$
for some $\gamma \in \bK(x, y)(\beta)$.
\end{problem}

In order to solve the above problem, we first present a vector version of the Hermite-like reduction in~\cite{GeddesLeLi2004}. Let $\vec{a} = \frac{1}{d}(a_1, \ldots, a_n)\in \bK(x, y)^n$ with $a_i, d \in \bK[x, y]$ satisfying that $\gcd(d, a_1, \ldots, a_n)=1$
and $\mb = \frac{1}{e}(b_{i, j})\in \bK(x, y)^{n \times n}$ with $e, b_{i, j}\in \bK[x, y]$ such that $\gcd(e, b_{1,1}, {\ldots}, b_{1, n}, {\ldots},  b_{n, n})=1$.
Let $p \in \bK[x, y]$ be any irreducible factor of $d$ that is coprime with $e$. Then $d = p^m d_1$ with $d_1\in \bK[x, y]$ and $\gcd(p, d_1)=1$.
Since $\gcd(p, D_y(p))=1$, we have $\gcd(p, D_y(p)d_1)=1$ and then the B\'ezout relation
\[ a_i = s_i p + t_i D_y(p)d_1,\]
where $s_i, t_i \in \bK(x)[y]$.
Using integration by parts, we get
\[ \frac{a_i}{p^m d_1}    = \frac{s_i p + t_i D_y(p)d_1}{p^m d_1}
= D_y \left(\frac{u_i}{p^{m-1}}\right) + \frac{v_i}{p^{m-1}d_1},  \]
where $u_i = t_i(1-m)^{-1} $ and $v_i = s_i - (1-m)^{-1} D_y(t_i) d_1$.
Let $\vec{u} = (u_1, \ldots, u_n)$ and $\vec{v} = (v_1, \ldots, v_n)$.
Then we have
\[\vec{a} {=} D_y\left(\frac{\vec{u}}{p^{m-1}}\right) + \frac{\vec{v}}{p^{m-1}d_1} {=} D_y\left(\frac{\vec{u}}{p^{m-1}}\right) +  \frac{\vec{u}}{p^{m-1}} \cdot \mb  + \frac{\vec{w}}{p^{m-1}d_1 e},\]
where $\vec{w} \in \bK(x)[y]^n$. Repeating this process yields
\[\vec{a} = D_y\left(\frac{\vec{g}}{p^{m-1}}\right) + \frac{\vec{g}}{p^{m-1}} \cdot \mb + \frac{\vec{h}}{pd_1 e},\]
where $\vec{g}, \vec{h}\in \bK(x)[y]^n$. By reducing the multiplicity of each irreducible factor of $d$ that is coprime with $e$ in the above way, we obtain
the additive decomposition
\begin{equation}\label{EQ:SDDred3}
\vec{a} = D_y(\vec{b}) + \vec{b} \cdot \mb + \vec{r},
\end{equation}
where~$\vec{b} \in \bK(x, y)^n$ and $\vec{r}= \frac{1}{p c} (r_1, \ldots, r_n)$ with $r_i \in \bK(x)[y]$
and $p, c\in \bK[x, y]$ be such that $p$ is a squarefree polynomial and $\gcd(p, e)=1$ and each irreducible factor of $c$ divides $e$.
We call the above process
a \emph{vector Hermite reduction} of $\vec{a}$ with respect to $\mb$.

Let $\beta\in \overline{\bK(y)}$ and $n = [\bK(y, \beta):\bK(y)]$.
Assume that $\{ \beta_1, \ldots, \beta_n \}$ is a basis for $\bK(y, \beta)$
as a linear space over $\bK(y)$.
Since $D_y(\beta_i) \in \bK(y, \beta)$, we have
$D_y(\beta_i) = \frac{1}{e}\sum_{j=1}^n b_{j, i} \beta_j$ with $e, b_{j, i}\in \bK[y]$. Set $\mb = \frac{1}{e}(b_{i, j}) \in \bK(y)^{n \times n}$. Then
$ D_y(\vec{\beta}) = \vec{\beta}\cdot \mb$ with $\vec{\beta} = (\beta_1, \ldots, \beta_n)$.
Since $\alpha  \in \bK(x, y)(\beta)$, we can write
$\alpha = \vec{a} \cdot \vec{\beta}^T$ for some $\vec{a} = \frac{1}{d}(a_1, \ldots, a_n)\in \bK(x, y)^n$ with $d, a_i \in \bK[x, y]$.
Applying the vector Hermite reduction to $\vec{a}$ with respect to $\mb$ yields the additive decomposition~\eqref{EQ:SDDred3}, which is equivalent to
\begin{equation}\label{EQ:reduceAlpha}
  \alpha = D_y(\vec{b} \cdot \vec{\beta}^T) + \tilde{\alpha} \, \, \text{with}\, \, \tilde{\alpha} = \frac{1}{pc} \sum_{i=1}^n r_i \beta_i,
\end{equation}
where $r_i, p, c\in \bK[x, y]$ with $p$ being squarefree and $\gcd(p, e)=1$ and each irreducible factor of $c$ divides $e\in \bK[y]$.

\begin{theorem}\label{THM:SDDsimple}
Let $\alpha\in \bK(x, y)(\beta)$ be of the form~\eqref{EQ:reduceAlpha}.
Then $\alpha$ has a telescoper
of type $(\partial_x, D_y)$ if and only if the polynomial $p$ in~\eqref{EQ:reduceAlpha} is split in $x$ and $y$.
\end{theorem}
\begin{proof}
Assume that $p$ is split in $x$ and $y$, i.e., $p = p_1p_2$ for some $p_1\in \bK[x]$ and $p_2\in \bK[y]$.
Then $\tilde{\alpha}$ can be written as $\tilde{\alpha} = \sum_{j=1}^m f_j \cdot g_j$ with $f_j \in \bK(x)$ and $g_j \in \bK(y)(\beta)$
since $\beta_i\in \bK(y)(\beta)$ and $c\in \bK[y]$. Let $L_j = f_j(x)\partial_x - \theta_x(f_j)\in \bK(x)\la \partial_x \ra$ for each $1\leq j\leq m$. Then $L_j(f_j \cdot g_j) = 0$.
So the LCLM of the $L_j$'s annihilates $\tilde{\alpha}$, which then is a telescoper for $\alpha$ of type $(\partial_x, D_y)$.
To show the necessity, we assume that $L = \sum_{\ell=0}^\rho e_{\ell} \partial_x^\ell\in\bK(x)\la\pa_x\ra$ with $e_0e_\rho \neq 0$ is a telescoper
for $\alpha$ of type $(\partial_x, D_y)$.
Then $L(\tilde{\alpha}) = D_y(\tilde{\gamma})$ for some $\tilde{\gamma}\in \bK(x, y)(\beta)$.
Write $\tilde{\gamma} = \vec{s} \cdot \vec{\beta}^T$ with $\vec{s} \in \bK(x, y)^n$
and $\vec{r} = (r_1, \ldots, r_n)$.
Then we have
\[ L\left(\frac{1}{pc} \vec{r}\right) = \sum_{\ell=0}^\rho \frac{e_\ell}{\theta_x^\ell(p)c} \theta_x^\ell(\vec{r})=D_y(\vec{s}) + \vec{s}\cdot \mb. \]
Suppose that $p$ is not split in $x$ and $y$. Then there exists a non-split irreducible factor $p_0$ of $p$ such that $\theta_x(p_0)\nmid p$.
Then $\theta_x^\rho(p_0)$ is also a non-split irreducible polynomial and only divides the denominator $\theta_x^{\rho}(p)c$.
Since $p$ is squarefree, the valuation of the left-hand side of the
above equality at $\theta_x^\rho(p_0)$ is $-1$. However, the valuation of the right-hand side is either $\geq 0$ or~$< -1$ since $\mb\in \bK(y)^{n\times n}$. This
leads to a contradiction. So $p$ is split in $x$ and $y$.
\end{proof}

\begin{example}\label{EXAM:SDD}
Let $f = x/(z^2-y)$. Then
\[f = \frac{\alpha}{z-\beta}+ \frac{ - \alpha}{z+\beta},\]
where $\alpha = x/(2\sqrt{y})$ and $\beta = \sqrt{y}$. By Theorem~\ref{THM:SDDcriteria}, $f$ has a telescoper of type~$(\partial_x, D_y, D_z)$
since $\beta\in \overline{\bK(y)}$ and $L = x\partial_x - \theta_x(x)$ is a telescoper for $\alpha$ of type $(\partial_x, D_y)$.
Indeed, $L$ is also a telescoper for $f$ of type $(\partial_x, D_y, D_z)$.
\end{example}

\subsection{Existence problems of fourth class }\label{SUBSECT:class4}

We continue to address the fourth class of the existence problems of telescopers for
rational functions in three variables. There are four cases in this class.
\begin{problem}\label{PROB:SSD}
Let $\partial_x \in \{S_x, T_{q, x}\}$ and $\Theta_y \in \{\Delta_y, \Delta_{q, y}\}$.
Given $f\in \bK(x, y, z)$, decide whether there exists a nonzero operator $L\in \bK(x)\langle \partial_x \rangle$
such that $L(f) = \Theta_y(g) + D_z(h)$ for some $g, h\in \bK(x, y, z)$.
\end{problem}

Let $(\pa_v,\theta_v)\in\{(S_v,\sigma_v),(T_{q,v},\tau_{q,v})\}$ for $v\in\{x,y\}$.
By the Ostrogradsky--Hermite reduction in $z$ and the reduction formula~\eqref{eq:red} with $\si = \theta_y$, we can decompose
$f$ as
\begin{equation}\label{EQ:redSSD}
f = \Theta_y(u) + D_z(v) + r, \,\,  \text{where~$r= \sum_{i=1}^I \sum_{j=0}^{J_i}\frac{a_{i, j}}{\theta_x^j(d_i)}$}
\end{equation}
with $a_{i, j}\in \bK(x, y)[z]$ and $d_i\in \bK[x, y, z]$ satisfying the condition $\deg_z(a_{i, j})< \deg_z(d_i)$
and the $d_i$'s are irreducible polynomials in distinct $\la \theta_x, \theta_y \ra $-orbits.
Note that $f$ has a telescoper of type $(\partial_x, \Theta_y, D_z)$
if and only if $r$ does.

\begin{lemma}\label{LEM:SSDred1}
Let $r {\in} \bK(x, y, z)$ be as in~\eqref{EQ:redSSD}. Then $r$ has a telescoper of type $(\partial_x, \Theta_y, D_z)$
if and only if for each $i$ with $1\leq i \leq I$, we have $r_i = \sum_{j=0}^{J_i}\frac{a_{i, j}}{\theta_x^j(d_i)}$
has a telescoper of the same type.
\end{lemma}
\begin{proof}
The sufficiency follows from Lemma~\ref{LM:closure}. For the necessity we assume that $L = \sum_{k=0}^\rho \ell_k \partial_x^k\in \bK(x)\la \partial_x\ra$ with $\partial_x \in \{S_x, T_{q, x}\}$ and $\ell_0 \neq 0$
is a telescoper for $r$ of type $(\partial_x, \Theta_y, D_z)$. Then
\[L(r) = \sum_{i=1}^I L(r_i) = \sum_{i=1}^I \left(\sum_{j=0}^{J_i + \rho}\frac{\sum_{k=0}^j \ell_k \theta_x^k(a_{i, j-k})}{\theta_x^j(d_i)}\right)\]
with $\ell_k = 0$ if $k>\rho$ and $a_{i, j}=0$ if $j>J_i$ is $(\Theta_y, D_z)$-exact.
Since the $d_i$'s are in distinct $\la \theta_x, \theta_y\ra$-orbits, the $\theta_x^j(d_i)$'s are
in distinct $\la \theta_y \ra$-orbits. By Lemma~\ref{LEM:SDExact},  we have $L(r_i)$ is $(\Theta_y, D_z)$-exact for each $i$ with $1\leq i \leq I$. So each $r_i$ has a telescoper of the same type.
\end{proof}

Now the existence problem is reduced to that for rational functions of the form
\begin{equation} \label{EQ:SSDred2}
f = \sum_{i=0}^I \frac{a_i}{\theta_x^i(d)},\end{equation}
where $a_i\in \bK(x, y)[z], d\in \bK[x, y, z]$ with $\deg_z(a_i)<\deg_z(d)$ and $d$ is irreducible in $z$ over $\bK(x, y)$.
We will proceed by a case distinction according to whether or not $d$ satisfies the condition:
there exist $c\in \bK\setminus\{0\}$ and  integers $m, n$ with $m > 0$ such that
\begin{equation}\label{EQ:SSDc1}
  \theta_x^{m}(d)= c \cdot \theta_y^{n}(d).
\end{equation}
Note that the constant $c$ in~\eqref{EQ:SSDc1} must be 1
if $(\theta_x, \theta_y) \in \{ (\si_x, \si_y), (\si_x, \tau_{q, y}), (\tau_{q, x}, \si_{y})\}$ by the comparison of leading coefficients.
When $(\theta_x, \theta_y) = (\tau_{q, x}, \tau_{q, y})$, we claim that $c = q^s$ for some $s\in \bZ$. To show this claim, we write
$d = \sum_{i, j, k} c_{i, j, k}x^iy^jz^k$. Then the equality $\tau_{q, x}^m(d) = c \tau_{q, y}^n(d)$
implies that for all $i, j$, we have $c = q^{im-jn}$. Let $s = \gcd(m, n)$. Then $m = s\bar m$ and $n= s \bar n$. For different pairs $(i_1, j_1)$ and $(i_2, j_2)$
with $q^{i_1m-j_1n} = q^{i_2m-j_2n}$, we have $i_1m-j_1n = i_2m-j_2n$ since $q$ is not a root of unity, which further implies that
$(i_2, j_2) = (i_1, j_1) + \lambda (\bar n, \bar m)$ for some nonzero $\lambda \in \bZ$. Thus $d = x^{i_0}y^{j_0} \bar d$, where $i_0, j_0 \in \bZ$ and $\bar d = \sum_{k=0}^\rho d_k(x^{\bar n}y^{\bar m})z^k$ with $d_k \in \bK[T]$. Since $\tau_{q, x}^m(\bar d) = \tau_{q, y}^n(\bar d)$, we have $c = q^{i_0m - j_0n}$.
Combing the above discussions with~\cite[Proposition 1]{chen2019wz} yields a characterization of polynomials satisfying the condition~\eqref{EQ:SSDc1}.
\begin{lemma} \label{LEM:invd}
Let $d= \sum_{i=0}^{\rho} d_i z^i \in \bK(x, y)[z]$ be a polynomial in $z$ over $\bK(x, y)$.
If there exist $c\in \bK\setminus\{0\}$ and $m, n\in \bZ$ with $m>0$ such that $\theta_x^m(d)=c \cdot \theta_y^n(d)$, then for each $i$ with
$0\leq i \leq \rho$ we have
\medskip
\begin{enumerate}
\item if $(\theta_x, \theta_y) = (\si_x, \si_y)$, then $c=1$ and $d_i$ is integer-linear in $x$ and $y$, i.e., $d_i = f(nx+my)$ for some $f \in \bK(z)$;
\item if $(\theta_x, \theta_y) = (\si_x, \tau_{q, y})$ or $ (\tau_{q, x}, \si_{y})$, then $c =1$ and $d_i \in \bK(y)$ and $d_i \in \bK$ if $n\neq 0$;
\item if $(\theta_x, \theta_y) = (\tau_{q, x}, \tau_{q, y})$, then $c = q^s$ for some $s\in \bZ$ and $d_i$ is $q$-integer-linear in $x$ and $y$, i.e., $d_i = x^{n_0}y^{m_0}f_i(x^ny^m)$ for some $f_i \in \bK(z)$ and $n_0, m_0 \in \bZ$.
\end{enumerate}
\end{lemma}

By the above characterization, the condition~\eqref{EQ:SSDc1} can be checked by solving the bivariate case of
Problems~\ref{PROB:shiftequiv} and~\ref{PROB:qshiftequiv} in the pure shift and $q$-shift cases, respectively.

\begin{lemma} \label{LEM:d1}
Let $f\in \bK(x, y, z)$ be of the form~\eqref{EQ:SSDred2} and $d$ does not satisfy the condition~\eqref{EQ:SSDc1}. Then
$f$ has a telescoper of type $(\partial_x, \Theta_y, D_z)$
if and only if $f$ is $(\Theta_y, D_z)$-exact.
\end{lemma}
\begin{proof} The sufficiency is clear by definition.
Assume that $L = \sum_{k=0}^\rho \ell_k \partial_x^k$ with $\ell_0 \neq 0$
is a telescoper for $f$ of type $(\partial_x, \Theta_y, D_z)$. Then we have that
\[L(f) = \sum_{i=0}^{\rho + I} \left( \frac{\sum_{j=0}^i \ell_j \theta_x^j(a_{i-j})}{\theta_x^i(d)}\right)\]
is $(\Delta_y, D_z)$-exact, where~$\ell_j=0$ if $j>\rho$ and $a_i=0$ if $i>I$.
Since $d$ does not satisfy the condition~\eqref{EQ:SSDc1}, we have $\theta_x^i(d)$ and $\theta_x^{i'}(d)$ in distinct $\la \theta_y\ra$-orbits for all $i\neq i'$.
By Lemma~\ref{LEM:SDExact}, for any $i$ with $0\leq i \leq \rho+I$, there exist $u_i, v_i\in \bK(x, y, z)$ such that
\begin{equation}\label{EQ:exact-i}
\frac{\sum_{j=0}^i \ell_j \theta_x^j(a_{i-j})}{\theta_x^i(d)} = \Theta_y(u_i) + D_z(v_i).
\end{equation}
To show that all fractions $a_i/\theta_x^i(d)$ are $(\Theta_y, D_z)$-exact, we proceed by induction.
The assertion is true for $i=0$ since $a_0/d = \Theta_y(u_0/\ell_0) + D_z(v_0/\ell_0)$. Suppose that we have
shown that $a_i/\theta_x^i(d)$ is $(\Theta_y, D_z)$-exact for $i=0, \ldots, s-1$ with $s\leq I$. By the equality~\eqref{EQ:exact-i} with $i=s $,
we get
\[ \frac{a_s}{\theta_x^s(d)} = \Theta_y\left(\frac{u_s}{\ell_0}\right) + D_z\left(\frac{v_s}{\ell_0}\right) - \sum_{j=1}^s \frac{\ell_j}{\ell_0} \theta_x^j\left(\frac{a_{s-j}}{\theta_x^{s-j}(d)}\right). \]
By the commutativity between $\theta_x$ and $\theta_y, \delta_z$ and Lemma~\ref{LEM:SDExact}, we have $a/\theta_x^i(d)$ is $(\Theta_y, D_z)$-exact
for any $i\in \bN$ if $a/d$ is.
By the induction hypothesis, we have  $\frac{\ell_j}{\ell_0} \theta_x^j({a_{s-j}}/{\theta_x^{s-j}(d)})$ is $(\Theta_y, D_z)$-exact
for all $1\leq j \leq s$. So are $a_s/\theta_x^s(d)$ and $f$.
\end{proof}

We now deal with the case in which $d$ satisfies the condition~\eqref{EQ:SSDc1}.
From now on, we will always assume that $m$ is the smallest positive integer such
that $\theta_x^m(d) = c \cdot \theta_y^n(d)$ for some $n\in \bZ$ and $c\in \bK\setminus\{0\}$.
By the reduction formula~\eqref{eq:red} with $\sigma =\theta_y$, the existence problem is further reduced to
that for rational functions of the form
\begin{equation} \label{EQ:SSDred3}
f =  \sum_{i=0}^{m-1} \frac{a_i}{\theta_x^i(d)},
\end{equation}
where $a_i\in \bK(x, y)[z], d\in \bK[x, y, z]$
with $\deg_z(a_i)<\deg_z(d)$ and $d$ is irreducible in $z$ over $\bK(x, y)$.

The following lemma is similar to Lemma 5.3 in~\cite{Chen2016}.
\begin{lemma} \label{LEM:d1}
Let $f\in \bK(x, y, z)$ be of the form~\eqref{EQ:SSDred3} and $d$ satisfy the condition~\eqref{EQ:SSDc1}. Then
$f$ has a telescoper of type $(\partial_x, \Theta_y, D_z)$
if and only if for each $i$ with $0\leq i \leq I$, the fraction $a_i/\theta_x^i(d)$ has a telescoper of the same type.
\end{lemma}
\begin{proof}
The sufficiency follows from Lemma~\ref{LM:closure}.
For the necessity direction, one can adapt the second part of the proof of~\cite[Lemma 5.3]{Chen2016} to the setting of telescopers of type $(\partial_x, \Theta_y, D_z)$
literally by interpreting $\equiv_{y, z} 0$ as being $(\Theta_y, D_z)$-exact.
\end{proof}

The above lemma further reduces the existence problem to that for simple fractions of the form
\begin{equation}\label{EQ:SSDred4}
f = \frac{a}{bd},
\end{equation}
where $a, d\in \bK[x, y, z], b\in \bK[x, y]$ satisfy that $\gcd(a, bd)=1$ and $\deg_z(a)<\deg_z(d)$, and
$d$ is irreducible and satisfies the condition~\eqref{EQ:SSDc1}.
We will consider two cases according to whether $d$ is in $\bK[x, z]$ or not.
If $d\in \bK[x, z]$, then $\theta_y^i(d)=d$ for all $i\in \bN$.
The condition $\theta_x^m(d) = \theta_y^n(d)$ implies that $d$ is also free of $x$,
i.e., $d\in \bK[z]$. Thus $L \in \bK(x)\la \partial_x\ra$ is a telescoper for $f$ of type
$(\partial_x, \Theta_y, D_z)$ if and only if $L(a/b)=\Theta_y(u)$ for some $u\in \bK(x, y)[z]$
with $\deg_z(u)< \deg_z(d)$. Write $a = \sum_{i=0}^{\deg_z(d)-1} a_i z^i$
and $u = \sum_{i=0}^{\deg_z(d)-1} u_i z^i$. Then for each $i$ with $0\leq i \leq  \deg_z(d)-1$, we have
$L(a_i/b) = \Theta_y(u_i)$, i.e., $L$ is a telescoper for all $a_i/b$ of type $(\partial_x, \Theta_y)$.
The existence problem is then reduced to that in the bivariate case, for which Theorem~\ref{THM:bicriteria} applies.
So it remains to deal with the case when $d$ is not in $\bK[x, z]$.

\begin{lemma}\label{LEM:torsion} Let $\tau := \theta_x^m\theta_y^{-n}$ with $m, n \in \bZ$ and $m>0$ and let $p \in \bK[x, y]$ be an irreducible polynomial.
If $\tau^i(p) = \lambda \cdot p$ for some nonzero $i\in \bZ$ and nonzero $\lambda\in  \bK$, then $\tau(p)=\mu \cdot p$ for some
 nonzero $\mu \in \bK$.
\end{lemma}
\begin{proof}
We prove by cases. Write $p = \sum_{i, j} p_{i, j} x^iy^j$ with $p_{i, j} \in \bK$. If $(\theta_x, \theta_y) = (\si_x, \si_y)$, then $\tau^i(p) = \lambda \cdot p$
implies that $\lambda = 1$ by comparing the leading coefficients. So $\si_x^{im}(p) = \si_y^{in}(p)$. By Lemma~\ref{LEM:invd}, we have $p = r(inx + imy)$
for some $r = \sum_{j=0}^s r_j z^j\in \bK[z]$. Thus $p = \tilde{r}(nx+my)$ with $\tilde{r} = \sum_{j=0}^s r_j i^j z^j$, which implies that $\tau(p) = p$.
If $(\theta_x, \theta_y) = (\si_x, \tau_{q, y})$, then $\tau^i(p) = \lambda \cdot p$
implies that $p \in \bK[y]$ and moreover $p = c\cdot y$ for some $c\in \bK$ if $n\neq 0$ by~\cite[Lemma 5.4]{Chen2015}, which leads to that $\tau(p) =\mu \cdot p$ with $\mu = q$.
If $(\theta_x, \theta_y) = (\tau_{q, x}, \si_{y})$, then $\tau^i(p) = \lambda \cdot p$
implies that $p \in \bK[y]$ and moreover $p \in \bK$ if $n\neq 0$ by~\cite[Lemma 5.4]{Chen2015}.
Then we have $\tau(p) =p$. If $(\theta_x, \theta_y) = (\tau_{q, x}, \tau_{q, y})$, then $\tau^i(p) = \lambda \cdot p$
implies that $p = (x^sy^t) \cdot r(x^{in}y^{im})$ for some $s, t\in \bZ$ and $r\in \bK[z]$ by~\cite[Lemma 5.2]{DuLi2019}.
So we have $\tau(p) = \mu \cdot  p$ with $\mu = q^{sm-nt}$.  This completes the proof.
\end{proof}

\begin{lemma}\label{LEM:split} Let $\tau := \theta_x^m\theta_y^{-n}$ with $m, n \in \bZ$ and $m>0$ and let $f = a/b$ with $ a, b \in \bK[x, y]$ and $\gcd(a, b)=1$.
 If there exist $e_0, \ldots, e_r\in \bK(x)$, not all zero, such that $\sum_{i=0}^r e_i \tau^i(f)=0$, then
 $b = b_1 b_2$ with $b_1\in \bK[x]$ and $b_2\in \bK[x, y]$ satisfying that $\tau(b_2) = \lambda \cdot b_2$ for some nonzero $\lambda \in \bK$.
\end{lemma}
\begin{proof} Assume that $\sum_{i=0}^r e_i \tau^i(f)=0$.
Let $b_1$ and $b_2$ be the content and primitive part of $b$ as a polynomial in $y$ over $\bK[x]$.
If $b_2$ is a constant in $\bK$, then the assertion holds since $\tau(b_2) = b_2$.
We now assume that $b_2\notin \bK$.
Then all of its irreducible factors have positive degree in $y$.
Assume that there exists an irreducible factor $p$ of $b_2$ such that $\tau(p)\neq c \cdot p$ for any $c\in \bK$.
Then for any integer $i\neq 0$, $\tau^i(p)\neq c_i \cdot  p$ for any $c_i \in \bK$ by Lemma~\ref{LEM:torsion}.
Among all of such irreducible factors, we can always find one factor
$p$ of multiplicity $m$ such that $\tau^i(p)\nmid b_2$ for all integer $i<0$.
Then $\tau^i(p)$ is also irreducible for all $i\in \bZ$ and $\gcd(\tau^i(p), \tau^j(p))=1$
if $i\neq j$. Let $s$ be the largest integer such that $\tau^s(p)\mid b_2$. Then the irreducible polynomial $\tau^{r+s}(p)$
only divides the denominator $\tau^r(b)$ and not others, which implies that $\sum_{i=0}^r e_i \tau^i(f)\neq 0$ since $p$ depends on $y$ and the coefficients $e_i$ are in $\bK(x)$. This leads to a contradiction. So for each irreducible factor $p$ of $b_2$ we have $\tau(p) =  c \cdot p$ for  some $c\in \bK$.
This implies that $\tau(b_2)=\lambda \cdot b_2$ for some $\lambda \in \bK$.
\end{proof}

\begin{lemma}\label{LEM:SSDsuff}
Let $a\in \bK(x)[y, z]$ and~$b\in \bK[x, y, z]$ be such that $b\neq 0$ and $\theta_x^m(b)=c \cdot \theta_y^n(b)$ for some $c\in \bK\setminus\{0\}$ and $m, n\in\bZ$ with $m>0$.
Then $a/b$ has a telescoper of type $(\partial_x, \Theta_y, D_z)$.
\end{lemma}
\begin{proof} Set $f = a/b$. It suffices to show that for sufficiently large $I\in \bN$,
there exist $\ell_0, \ldots, \ell_I\in \bK(x)$, not all zero, and $g\in \bK(x, y, z)$ such that
$L(f) = \Theta_y(g)$ with $L=\sum_{i=0}^I \ell_i \partial_x^{im}$. By the reduction formula~\eqref{eq:red} with $\si =\theta_y$, we have
\[\theta_x^{im}(f) = \frac{\theta_x^{im}(a)}{\theta_x^{im}(b)} = \frac{\theta_x^{im}(a)}{c^i \cdot \theta_y^{in}(b)} = \Theta_y(g_i) + \frac{\theta_y^{-in}\theta_x^{im}(a)}{c^i\cdot b}\]
for some $g_i\in \bK(x, y, z)$. Note that the degrees of the polynomials $\theta_y^{-in}\theta_x^{im}(a)$ in $y$ and $z$ are the same as that of $a$.
So all the polynomials $\theta_y^{-in}\theta_x^{im}(a)$ lie in a finite dimensional linear space over $\bK(x)$. Therefore, for sufficiently large $I$,
there exist $\ell_0, \ldots, \ell_I\in \bK(x)$, not all zero, such that $\sum_{i=0}^I \ell_i \theta_y^{-in}\theta_x^{im}(a) = 0$. This implies that
$L$ is a telescoper for $f$ of type $(\partial_x, \Theta_y, D_z)$.
\end{proof}

\begin{theorem}\label{THM:criteriaSSD}
Let $f\in \bK(x, y, z)$ be of the form~\eqref{EQ:SSDred4}.  Assume that
$d$ is not in $\bK[x, z]$.
Then $f$ has a telescoper of type $(\partial_x, \Theta_y, D_z)$
if and only if $b=b_1 b_2$ for some $b_1\in \bK[x]$ and $b_2\in \bK[x, y]$ satisfying $\theta_x^m(b_2) = \lambda \cdot \theta_y^n(b_2)$
for some nonzero $\lambda\in \bK$.
\end{theorem}
\begin{proof}
The sufficiency follows from Lemma~\ref{LEM:SSDsuff}. For the necessity, we assume that $L\in \bK(x)\la \partial_x \ra$
is a telescoper for $f$ of type $(\partial_x, \Theta_y, D_z)$. Write $L=L_0+L_1+\cdots+L_{m-1}$ with $L_i = \sum_{j=0}^{r_i} \ell_{i, j} \partial_x^{jm+i}$.
Since $\theta_x^i(d)$ and $\theta_x^j(d)$ are in distinct $\la\theta_y\ra$-orbits for all $0\leq i \neq j \leq m-1$, Lemma~\ref{LEM:SDExact} implies that
$L_i$ is also a telescoper for $f$ of the same type for each $i$ with $0\leq i \leq m-1$. A direct calculation yields
\[L_0(f) = \Theta_y(g_0) +  \frac{A}{d},\]
where $A= \sum_{j=0}^{r_0} c^{-j}\ell_{0, j}\tau^j(a/b)$ with $\tau = \theta_y^{-n}\theta_x^{m}$ and $\tau(d) = c\cdot  d$. By Lemma~\ref{LEM:SDExact}, we have $A=0$ since $d\notin \bK[x, z]$.
So the necessity follows from Lemma~\ref{LEM:split}.
\end{proof}

\begin{example}\label{EXAM:SSD}
Let $f = 1/(bd)$ with $b =x+y$ and $d = z^2-x-y$. Note that $d$ satisfies the condition~$\si_x(d) = \si_y(d)$ and is not in~$\bK[x, z]$.
By Theorem~\ref{THM:criteriaSSD}, $f$ has a telescoper of type $(S_x, \Delta_y, D_z)$ since $b$ satisfies the same condition as $d$.
Indeed, $L = S_x -1$ is a telescoper for $f$ since $L(f) = \Delta_y(f) + D_z(0)$.
\end{example}

\subsection{Existence problems of fifth class }\label{SUBSECT:class5}
We now consider the fifth class of existence problems in which both telescopers and
$(\Theta_y, \Theta_z)$ are involving  ($q$-)shift operators. In this class, we let
$\pa_x\in\{S_x,T_{q,z}\}$ and $(\Theta_y,\Theta_z)\in\{(\Delta_y,\Delta_z),
(\Delta_{q,y},\Delta_{z}),(\Delta_{q,y},\Delta_{q,z})\}$. More precisely, we
solve the following problem.

  \begin{problem}\label{Problem:type5}
Given $f\in\bK(x,y,z)$, determine if there exists a nonzero operator $L\in\bK(x)\la \pa_x \ra$ such that
$L(f)=\Theta_y(g)+\Theta_z(h)$ for some $g,h\in\bK(x,y,z)$.
\end{problem}
For $v\in\{x,y,z\}$, let $\theta_v=\si_v$ if $\Theta_v=\Delta_v$ or $\theta_v=\tau_{q,v}$ if $\Theta_v=\Delta_{q,v}$.
By partial fraction decomposition w.r.t $z$ and the transformation~\eqref{eq:red} with $\theta=\theta_y$ and subsequently with $\theta=\theta_z$, any rational function $f\in\bK(x,y,z)$ can be decomposed into
\begin{equation}\label{eq:reduced form3}
f=\Theta_y(u)+\Theta_z(v)+\mu+
\sum_{i=1}^{I}\sum_{j=1}^{J_i}\sum_{\ell=0}^{t_{i, j}}
\frac{a_{i,j,\ell}}{\theta_x^{\ell}d_i^j},
\end{equation}
where $u,v\in\bK(x,y,z),\mu\in\bK(x,y),a_{i,j,\ell}\in\bK(x,y)[z],d_i\in\bK[x,y,z]$ with $\deg_z(a_{i,j,\ell})<\deg_z(d_i)$, $d_i$'s are
irreducible polynomials in distinct $\la\theta_x,\theta_y,\theta_z\ra$-orbits, $\theta_x^{\ell}d_i$ and $\theta_x^{\ell'}d_i$
are not $\la\theta_y,\theta_z\ra$-equivalent for any $1\leq i\leq I$, $0\leq\ell, \ell'\leq t_{i, j}$ with $\ell\neq\ell'$.
%
Then by similar discussions as the proof of Lemma 5.2 and Lemma 5.3 in~\cite{Chen2016}, we can obtain the following result.
\begin{lemma}
Let $f\in\bK(x,y,z)$ be of the form \eqref{eq:reduced form3}.
Then $f$ has telescopers of type $(\pa_x,\Theta_y,\Theta_z)$ if and only if
$\mu$ and all $\frac{a_{i,j,\ell}}{\theta_x^{\ell}d_i^j}$ with $1\leq i\leq I,1\leq j\leq J_i$ and $0\leq \ell \leq t_{i, j}$ have telescopers of the same type.
\end{lemma}
Notice that for $\mu\in\bK(x,y)$, having telescopers of type $(\pa_x,\Theta_y,\Theta_z)$ and $(\pa_x,\Theta_y)$ are equivalent.
The existence problem of bivariate rational functions has been solved by Theorem \ref{THM:bicriteria}.
Thus Problem~\ref{Problem:type5} for a general rational function has been reduced to that for a rational function of the form
\begin{equation}\label{eq:simple fraction5}
f=\frac{b(x,y,z)}{c(x,y)d(x,y,z)^{\lambda}},
\end{equation}
where $\lambda\in\bN\setminus\{0\}$, $c\in\bK[x,y]$, $b,d\in\bK[x,y,z]$ with $0\leq\deg_z(b)<\deg_z(d)$.
Suppose $\alpha(x)\in\bK(x)\setminus\{0\}$.
It is easy to check that
$$
\sum_{i=1}^{\rho}a_i(x)\pa_x^i(\alpha f)
=\sum_{i=1}^{\rho}\left(a_i(x)\pa_x^i(\alpha)\right)\pa_x^i(f)
$$
whenever $a_i(x)\in\bK(x)$ and $f\in\bK(x,y,z)$.
This means the existence problem of $f$ is equivalent to that of $\alpha f$.
As such we can assume in the form \eqref{eq:simple fraction5} that $b,c,d$ are all primitive in $y,z$.
If $f$ is $(\Theta_y,\Theta_z)$-exact.
Then $L=1$ is a telescoper for $f$.
From now on, we will also assume $f$ is not $(\Theta_y,\Theta_z)$-exact.
\begin{lemma}
Let $f\in\bK(x,y,z)$ be of the form \eqref{eq:simple fraction5}.
If $f$ has a telescoper of type $(\pa_x,\Theta_y,\Theta_z)$ then
\begin{equation}\label{eq:condition of d}
\theta_x^m(d)=q^s\theta_y^n\theta_z^k(d)\quad \text{ for some } m,s,n,k\in\bZ \text{ with } m>0.
\end{equation}
\end{lemma}
\proof
We prove the claim by contradiction.
Suppose the condition \eqref{eq:condition of d} does not hold.
Assume that $L=\sum_{i=0}^{I}a_i\pa_x^i\in\bK(x)\la\pa_x\ra$ with $a_0\neq0$ is a telescoper for $f$.
Then
\[
L(f)=\sum_{i=0}^{I}\frac{a_i\theta_x^i(b)}{\theta_x^i(c)\theta_x^{i}(d^{\lambda})}
=\Theta_y(g)+\Theta_z(h)
\]
for some $g,h\in\bK(x,y,z)$.
By assumption, we know $\theta_x^i{d}$'s are in distinct $\la\theta_y,\theta_z\ra$-orbits, Lemmas \ref{LEM:SSExact},~\ref{LEM:TTExact} and \ref{LEM:STExact} show that
for any $0\leq i\leq I$, $\frac{a_i\theta_x^i(b)}{\theta_x^i(c)\theta_x^{i}(d^{\lambda})}$ are $(\Theta_y,\Theta_z)$-exact.
Particularly,
\[
\frac{a_0b}{cd^{\lambda}}=\Theta_y(g_0)+\Theta_z(h_0)\text{ for some }g_0,h_0\in\bK(x,y,z).
 \]
As $a_0\in\bK(x)\setminus\{0\}$, we get $\frac{b}{cd^{\lambda}}=\Theta_y(\frac{g_0}{a_0})+\Theta_z(\frac{h_0}{a_0})$ which contradicts to the assumption that $f$ is not $(\Theta_y,\Theta_z)$-exact.
This completes the proof.
\qed

Next, we will proceed by case distinction according to whether or not
\begin{equation}\label{eq:2condition of d}
\theta_y^{n_1}(d)=q^{s_1}\theta_z^{k_1}(d) \quad \text{ for some } s_1,n_1,k_1\in\bZ \text{ with } n_1>0.
\end{equation}
\begin{theorem}\label{th:ThtC}
Let $f\in\bK(x,y,z)$ be of the form \eqref{eq:simple fraction5} and $d$ satisfy the condition \eqref{eq:condition of d} but not the condition \eqref{eq:2condition of d}.
Then $f$ has a telescoper of type $(\pa_x,\Theta_y,\Theta_z)$ if and only if
\begin{equation}\label{eq:condition of c}
\theta_x^{tm}(c)=q^{s_2}\theta_y^{tn}(c)
\end{equation}
for the $(m,n)$ as in \eqref{eq:condition of d} and some $t,s_2\in\bZ$ with $t>0$.
\end{theorem}
\proof
For the sufficiency, assume that $c$ satisfies the condition \eqref{eq:condition of c}.
Then set
$
L = \sum_{i=0}^{I}a_{i} \pa_x^{i t m},
$
where $I\in \bN$ and $a_{i} \in \bK(x)$ are to be determined.
Applying the reduction formula~\eqref{eq:red} yields
\begin{align*}
  L(f) = \sum_{i=0}^I \frac{a_i q^{-is_2-its} \theta_x^{itm}(b)}{\theta_y^{itn}(c)\theta_y^{itn}\theta_z^{itk}(d^\lambda)}
      = \Theta_y(u) + \Theta_z(v) + \frac{1}{cd^\lambda} \sum_{i=0}^I a_i q^{-is_2-its} \theta_x^{itm}\theta_y^{-itn}\theta_z^{-itk}(b)
\end{align*}
for some~$u, v\in \bK(x, y,z)$.
Note that the degrees of the polynomials $\theta_x^{itm}\theta_y^{-itn}\theta_z^{-itk}(b)$ in $y$ or $z$ are the same as that of~$b$.
Thus all shifts of~$b$ lie in a finite dimensional linear space over~$\bK(x)$.
If $I$ is large enough, then there always exist $a_i\in \bK(x)$, not all zero, such that
$
\sum_{i=0}^I a_i q^{-is_2-its} \theta_x^{itm}\theta_y^{-itn}\theta_z^{-itk}(b)=0.
$
As a result $L = \sum_{i=0}^I a_i \pa_x^{itm}$ is a telescoper for $f$.

For the necessity, assume $f=\frac{b(x,y,z)}{c(x,y)d(x,y,z)^{\lambda}}$ has a telescoper $L_1$ of type $(\pa_x,\Theta_y,\Theta_z)$.
Let $C_1$ be the maximal factor of $c$ satisfying the condition \eqref{eq:condition of c} and $C_2=c/C_1$.
If $C_2\in\bK$ then we have done.
Now assume that $C_2\not\in\bK$.
Then $\deg_y(C_2)>0$ since $c$ is primitive with respect to $y,z$.
It follows that there exist $B_1,B_2\in\bK[x,y,z]$ with $\deg_z(B_i)<\deg_z(d)$ and $\gcd(B_i,C_i)=1$ for $i=1,2$, such that
\begin{equation*}
 f=\frac{1}{d^{\lambda}}\left(\frac{B_1}{C_1}+\frac{B_2}{C_2}\right),
\end{equation*}
Then $\frac{B_1}{C_1d^{\lambda}}$ has a telescoper $L_2$ of type $(\pa_x,\Theta_y,\Theta_z)$ by the sufficiency.
The least common left multiple of $L_1$ and $L_2$ is a telescoper for $\frac{B_2}{C_2d^{\lambda}}$.
Since $d$ satisfies the condition \eqref{eq:condition of d}, we can assume
$L=\sum_{i=0}^{I}a_i\pa_x^{im}\in\bK(x)\la\pa_x\ra$ with $a_0a_{I}\neq 0$ to be a telescoper for $\frac{B_2}{C_2d^{\lambda}}$. Thus
\begin{equation}\label{eq:B2C2}
L\left(\frac{B_2}{C_2d^{\lambda}}\right)
=\sum_{i=0}^{I}\frac{q^{-is}a_i\theta_x^{im}(B_2)}
                    {\theta_x^{im}(C_2)\theta_y^{in}\theta_z^{ik}(d^{\lambda})}
=\Theta_y(u)+\Theta_z(v)
+\sum\limits_{i=0}^{I}
           \frac{q^{-is}a_i\theta_x^{im}\theta_y^{-in}\theta_z^{-ik}(B_2)}
                {\theta_x^{im}\theta_y^{-in}(C_2)d^{\lambda}}
\end{equation}
for some $u,v\in\bK(x,y,z)$.
Notice that $L\left(\frac{B_2}{C_2d^{\lambda}}\right)$ is $(\Theta_y,\Theta_z)$-exact and that $d$ does not satisfy the condition\eqref{eq:2condition of d}.
Then Lemma \ref{LEM:SSExact}, \ref{LEM:TTExact} and \ref{LEM:STExact}  lead to
\begin{equation}\label{eq:NumIs0}
\sum\limits_{i=0}^{I}\frac{q^{-is}a_i\theta_x^{im}\theta_y^{-in}\theta_z^{-ik}(B_2)}
                {\theta_x^{im}\theta_y^{-in}(C_2)}=0.
\end{equation}
Let $\Lambda=\{c_j\in\bK[x,y]\setminus\bK[x] |\ c_j \text{ is an irreducible factor of }C_2\}$.
Then $\Lambda$ is nonempty and finite and none of $c_j$ satisfies condition \eqref{eq:condition of c} by the maximality of $C_1$.
By the method of proof by contradiction, one can prove that there exists a $c_{\ell}\in\Lambda$ such that
$c_{\ell}\neq q^{s'}\theta_x^{im}\theta_y^{-in}c_j$ for any
$c_j\in\Lambda$ and $s',i\in\bZ$ with $i>0$.
This fact together with equation \eqref{eq:NumIs0} and the constraint $\gcd(B_2,C_2)=1$ derive $B_2=0$, which concludes the proof.
\qed


\begin{lemma}\label{existence-lemma}
Let $f\in\bK(x,y,z)$ be of the form~\eqref{eq:simple fraction5} and $d$ satisfy conditions \eqref{eq:condition of d} and \eqref{eq:2condition of d}.
Suppose
\begin{equation}\label{eq:c}
\theta_x^{m_2}(c)=q^{s_2}\theta_y^{n_2}(c) \quad \text{ for integers }m_2,s_2,n_2 \text{ with }m_2>0.
\end{equation}
Then $f$ has a telescoper of type $(\pa_x,\Theta_y,\Theta_z)$.
\end{lemma}
\begin{proof}
Since $d$ satisfies both \eqref{eq:condition of d} and \eqref{eq:2condition of d}, without lose of generality, we assume $m,n_1$ are the smallest positive integers.
Let $m_0=mm_2n_1$ and
$
L = \sum_{i=0}^{I}a_{i} \pa_x^{i  m_0},
$
where~$I\in \bN$ and~$a_{i} \in \bK(x)$ are to be determined.
Then
\begin{align}
L\left(f\right)
& =\sum_{i=0}^I\frac{a_i\theta_x^{im_0}(b)}
                    {\theta_x^{im_0}(c)\theta_x^{im_0}(d^{\lambda})}\nonumber =\sum_{i=0}^I\frac{a_i q^{-ims_2n_1-ism_2n_1}\theta_x^{im_0}(b)}
        {\theta_y^{imn_2n_1}(c)\theta_y^{inm_2n_1}
         \theta_z^{ikm_2n_1}(d^{\lambda})}\nonumber  =\sum_{i=0}^I\frac{a_i q^{\alpha}\theta_x^{im_0}(b)}
        {\theta_y^{imn_2n_1}\theta_z^{\beta}(cd^{\lambda})}\nonumber \\
        & =\Theta_y(u)+\Theta_z(v)+
    \frac{\sum_{i=0}^Ia_i q^{\alpha}\theta_x^{im_0}\theta_y^{-imn_2n_1}\theta_z^{-\beta}(b)}
        {cd^{\lambda}},\label{eq:B2/C2}
\end{align}
where $u,v\in\bK(x,y,z)$,
$\alpha=-ims_2n_1-ism_2n_1-i(m_2n - mn_2)s_1$
and
$\beta=ikm_2n_1+i(m_2n - mn_2)k_1.$
Since the ($q$-)shift operators do not change the degree of $b$,
when $I$ is large enough, we can find nontrivial solutions $a_i$ such that
\[
\sum_{i=0}^Ia_i q^{\alpha}\theta_x^{im_0}\theta_y^{-imn_2n_1}\theta_z^{-\beta}(b)=0.
\]
Then identity \eqref{eq:B2/C2} leads to the fact that $L = \sum_{i=0}^I a_i \pa_x^{im_0}$ is a telescoper for $f$.
\end{proof}

\begin{theorem}
Let $f$ be of the form \eqref{eq:simple fraction5} and assume that $d$ satisfies conditions \eqref{eq:condition of d} and \eqref{eq:2condition of d}.
Then $f$ has a telescoper of type $(\pa_x,\Theta_y,\Theta_z)$ if and only if $f$ can be decomposed into the form
\begin{equation*}
f=\frac{1}{d^{\lambda}}\left(\frac{B_1}{C_1}+\frac{B_2}{C_2}\right),
\end{equation*}
where $B_1,B_2\in\bK[x,y,z]$, $C_1,C_2\in\bK[x,y]$ satisfy the following two constrains: (1) $C_1$ satisfies the condition~\eqref{eq:c};
(2) ${B_2}/{(C_2d^{\lambda})}$
is $(\Theta_y,\Theta_z)$-exact.
\end{theorem}
\begin{proof}
The sufficiency follows from Lemma~\ref{existence-lemma}.
For the necessity, let $C_1$ be the maximal factor of $c$ satisfying the condition \eqref{eq:c} and $C_2=c/C_1$.
If $C_2\in\bK$ then we have done.
Now assume that $C_2\not\in\bK$.
Then $\deg_y(C_2)>0$ since $c$ is primitive with respect to $y,z$.
It follows that there exist $B_1,B_2\in\bK[x,y,z]$ with $\deg_z(B_i)<\deg_z(d)$ and $\gcd(B_i,C_i)=1$ for $i=1,2$, such that
$
  f=\frac{1}{d^{\lambda}}\left(\frac{B_1}{C_1}+\frac{B_2}{C_2}\right).
$
Next we will prove
$
\frac{B_2}{C_2d^{\lambda}}
$
is $(\Theta_y,\Theta_z)$-exact.
Note that $\frac{B_1}{C_1d^{\lambda}}$ has a telescoper of the same type with $f$ by Lemma~\ref{existence-lemma}.
Then $\frac{B_2}{C_2d^{\lambda}}$ has a telescoper $L=\sum_{i=0}^{I}a_i\pa_x^{im}$ and
\begin{equation}\label{eq:c2}
L\left(\frac{B_2}{C_2d^{\lambda}}\right)=\Theta_y(u)+\Theta_z(v)
+\sum\limits_{i=0}^{I}\frac{q^{-is}a_i\theta_x^{im}\theta_y^{-in}\theta_z^{-ik}(B_2)}
                {\theta_x^{im}\theta_y^{-in}(C_2)d^{\lambda}}
\end{equation}
for some $u,v\in\bK(x,y,z)$.
Since $\deg_z(B_2)<\deg_z(d)$,
function
$\sum\limits_{i=0}^{I}
           \frac{q^{-is}a_i\theta_x^{im}\theta_y^{-in}\theta_z^{-ik}(B_2)}
                {\theta_x^{im}\theta_y^{-in}(C_2)d^{\lambda}}$
is $(\Theta_y,\Theta_z)$-exact and $d$ satisfies condition \eqref{eq:2condition of d},
exactness criteria in Lemmas~\ref{LEM:SSExact},~\ref{LEM:TTExact} and~\ref{LEM:STExact} yield that
there exists $g\in\bK(x,y)[z]$ such that
\begin{equation}\label{eq:NumDiffEqu}
\sum\limits_{i=0}^{I}
q^{-is}a_i\theta_x^{im}\theta_y^{-in}\theta_z^{-ik}\left(\frac{B_2}{C_2}\right)
=q^{-\lambda s_1}\theta_y^{n_1}\theta_z^{-k_1}(g)-g.
\end{equation}

Let $\Lambda=\{c_j\in\bK[x,y]\setminus\bK[x]\ |\ c_j \text{ is an irreducible factor of }C_2\}$.
Then $\Lambda$ is nonempty and finite since $\deg_y(C_2)>0$.
Notice that none of $c_j$ in $\Lambda$ satisfies the condition \eqref{eq:c}.
One can find a $c_{\ell}\in\Lambda$ such that
$c_{\ell}\neq q^{s}\theta_x^{m_3}\theta_y^{n_3}c_j$ for any
$c_j\in\Lambda$ and $s,m_3,n_3\in\bZ$ with $m_3>0$.
Collecting all irreducible factors in $C_2$, which are $\la \theta_y \ra$-equivalent to $c_{\ell}$, into $D_1$.
Then we can decompose $\frac{B_2}{C_2}$ into
$
  \frac{B_2}{C_2}=\frac{A_1}{D_1}+\frac{A}{D},
$
where $A_1,A\in\bK[x,y,z],D=C_2/D_1$.
Rewrite $g=g_1+g_2$ where $g_1,g_2\in\bK(x,y)[z]$ and
the denominator of $g_1$ contains exactly all irreducible factors in the denominator of $g$ which are $\la\theta_y\ra$-equivalent to $c_{\ell}$.
Equation \eqref{eq:NumDiffEqu} and the choice of $D_1$ and $g_1$ derive
$
  \frac{A_1}{D_1}=q^{-\lambda s_1}\theta_y^{n_1}\theta_z^{-k_1}(g_1)-g_1,
$
and hence
\begin{equation}\label{eq:A1D1}
\sum\limits_{i=0}^{I}
q^{-is}a_i\theta_x^{im}\theta_y^{-in}\theta_z^{-ik}\left(\frac{A_1}{D_1}\right)
=q^{-\lambda s_1}\theta_y^{n_1}\theta_z^{-k_1}(h_1)-h_1,
\end{equation}
where $h_1=\sum\limits_{i=0}^{I}
q^{-is}a_i\theta_x^{im}\theta_y^{-in}\theta_z^{-ik}(g_1)$.
Subtracting Equation \eqref{eq:A1D1} from \eqref{eq:NumDiffEqu}, we obtain
\begin{equation}\label{eq:Eqn1}
\sum\limits_{i=0}^{I}
q^{-is}a_i\theta_x^{im}\theta_y^{-in}\theta_z^{-ik}\left(\frac{A}{D}\right)
=q^{-\lambda s_1}\theta_y^{n_1}\theta_z^{-k_1}(g_1^{\star})-g_1^{\star}
\end{equation}
with $g_1^{\star}=g-h_1$.
Repeating the above arguments for the equation \eqref{eq:Eqn1}, one can finally decompose
$\frac{B_2}{C_2}=\frac{A_1}{D_1}+\frac{A_2}{D_2}+\cdots+\frac{A_T}{D_T}$ for $D_i\in\bK[x,y]$ and
$\frac{A_i}{D_i}=q^{-\lambda s_1}\theta_y^{n_1}\theta_z^{-k_1}(g_i)-g_i$ for any $1\leq i\leq T$.
Then we get
\[
\frac{B_2}{C_2}=q^{-\lambda s_1}\theta_y^{n_1}\theta_z^{-k_1}\left(\sum_{i=0}^{T}g_i\right)
      -\sum_{i=0}^{T}g_i
\]
and hence $\frac{B_2}{C_2d^{\lambda}}$ is $(\Theta_y,\Theta_z)$-exact.
This completes the proof.
\end{proof}

\subsection{Existence problems of sixth class }\label{SUBSECT:class6}
We consider the last class of the existence problems of telescopers for
rational functions in three variables.

\begin{problem}\label{PROB:DSD}
Let $\partial_y \in \{S_y, T_{q, y}\}$ and $\Theta_y = \partial_y - 1$. Given $f\in \bK(x, y, z)$, decide whether there exists a nonzero operator $L\in \bK(x)\langle D_x \rangle$
such that $L(f) = \Theta_y(g) + D_z(h)$ for some $g, h\in \bK(x, y, z)$.
\end{problem}

By the Ostrogradsky--Hermite reduction and the reduction formula~\eqref{eq:red}, we can decompose
$f\in \bK(x, y, z)$ as
\begin{equation}\label{EQ:DSDred1}
f = \Theta_y(u) + D_z(v) + r \, \, \text{with}\, \, r = \sum_{i=1}^I \frac{\alpha_i}{z-\beta_i},
\end{equation}
where $u, v\in \bK(x, y, z)$ and~$\alpha_i, \beta_i\in \overline{\bK(x, y)}$ with $\alpha_i\neq 0$ and the $\beta_i$'s are in
distinct $\la \theta_y \ra$-orbits with $\theta_y\in \{\si_y, \tau_{q, y}\}$.
Then $f$ has a telescoper of type $(D_x, \Theta_y, D_z)$ if and only if $r$
has a telescoper of the same type.

\begin{lemma}\label{LEM:Lhr}
For any $L= \sum_{j=0}^\rho  \ell_j D_x^j \in \bK(x)\la D_x \ra$ and $\alpha, \beta\in \overline{\bK(x, y)}$, there exists
$g\in  \overline{\bK(x, y)}(z)$ such that
\begin{equation}\label{EQ:Lhr}
L\left(\frac{\alpha}{z-\beta}\right)=\frac{L(\alpha)}{z-\beta} + D_z(g).
\end{equation}
\end{lemma}
\begin{proof}
Let $\text{res}_z(f, \beta)$ denote the residue of $f\in\bK(x,y,z)$ at $z=\beta$ in $z$. The map $\text{res}_z(\cdot, \beta)$
is $\bK(x, y)$-linear and commutes with the operator $D_x$ by~\cite[Proposition 3]{Chen2012}. Then we have
\[\text{res}_z\left(L\left(\frac{\alpha}{z-\beta}\right), \beta\right)= L\left(\text{res}_z\left(\frac{\alpha}{z-\beta}, \beta\right)\right)= L(\alpha).\]
So all residues of~$h := L(\alpha/(z-\beta))-L(\alpha)/(z-\beta)$ at all of its poles are zero. By Proposition 2.2 in~\cite{ChenSinger2012}, we have
$h$ is $D_z$-exact, i.e., $h=D_z(g)$ for some $g\in \overline{\bK(x, y)}(z)$.
\end{proof}
The next theorem reduces Problem~\ref{PROB:DSD} to the separation problem for algebraic functions (Problem~\ref{PROB:sep})
and the existence problem of telescopers in $\bK(x, y)(\beta)$ with $\beta\in \overline{\bK(x)}$.

\begin{theorem}\label{THM:criteriaDSD}
Let $f\in \bK(x, y, z)$ be of the form~\eqref{EQ:DSDred1}.
Then $f$ has a telescoper of type $(D_x, \Theta_y, D_z)$ if and only if for each $i$ with $1\leq i \leq I$, either
$\alpha_i$ is separable in $x$ and $y$ or $\beta_i\in \overline{\bK(x)}$ and $\alpha_i \in \bK(x, y)(\beta_i)$ has a telescoper of type $(D_x, \Theta_y)$.
\end{theorem}
\begin{proof}
If for each $i$ with $1\leq i \leq I$, either $\alpha_i$ is separable or $\beta_i\in \overline{\bK(x)}$ and $\alpha_i \in \bK(x, y)(\beta_i)$
has a telescoper of type $(D_x, \Theta_y)$,
then there exists a nonzero $L_i\in \bK(x)\la D_x\ra$ such that either $L_i(\alpha_i) = 0$ or $L_i(\alpha_i) = \Theta_y(\gamma_i)$ for some $\gamma_i\in \bK(x, y)(\beta_i)$.
By Lemma~\ref{LEM:Lhr}, we have
\begin{align*}
  L_i \left(\frac{\alpha_i}{z-\beta_i}\right) &  = D_z(g_i) + \frac{L_i(\alpha_i)}{z-\beta_i} = D_z(g_i) + \frac{\Theta_y(\gamma_i)}{z-\beta_i}\\
   & = D_z(g_i) + \Theta_y \left(\frac{\gamma_i}{z-\beta_i}\right),
\end{align*}
where $g_i \in \overline{\bK(x, y)}(z)$. So for each $i$ with $1\leq i \leq I$, the fraction $\alpha_i/(z-\beta_i)$
has a telescoper of type $(D_x, \Theta_y, D_z)$. Then $f$ has a telescoper of the same type by
Lemmas~\ref{LM:closure} and~\ref{LEM:trace}. To show the necessity, we assume that $L\in \bK(x)\la D_x \ra$ is a  telescoper for $f$ of type $(D_x, \Theta_y, D_z)$.
By Lemma~\ref{LEM:Lhr}, there exists $w\in \overline{\bK(x, y)}(z)$ such that
\begin{align*}
L(f)   &  = \Theta_y(L(u)) + D_z(L(v) + w) + \sum_{i=1}^I \frac{L(\alpha_i)}{z-\beta_i} \\
   & =\Theta_y(g) + D_z(h)
\end{align*}
for some $g, h \in \bK(x, y, z)$. For each $i$ with $1\leq i \leq I$, either $\alpha_i$ is separable if $L(\alpha_i)=0$ or
$L(\alpha_i)/(z-\beta_i)$ is $(\Theta_y, D_z)$-exact if $L(\alpha_i) \neq 0$. In the later case we have
$\beta_i\in \overline{\bK(x)}$ and $L(\alpha_i)=\Theta_y(\gamma_i)$ for some $\gamma_i\in \bK(x, y)(\beta_i)$ by Lemma~\ref{LEM:SDExact}.
\end{proof}

\begin{remark} The separation problem on algebraic functions will be solved in the forthcoming paper~\cite{Chen2019}.
The existence problem of telescopers of type $(D_x, \Theta_y)$ can be verified by Theorem \ref{THM:bicriteria}, whose statement is for functions in $\bK(x, y)$,
but its proof also works for functions in~$\overline{\bK(x)}(y)$.
In particular, this covers the case in which the functions are in $\bK(x, y)(\beta)$ with $\beta\in \overline{\bK(x)}$.
\end{remark}

\begin{example}\label{EXAM:DSD}
Let $f$ be as in Example~\ref{EXAM:SSD}. Then
\[f = \frac{\alpha}{z-\beta} + \frac{-\alpha}{z+\beta},\]
where $\alpha = \frac{1}{2(x+y)\sqrt{x+y}}$ and $\beta = \sqrt{x+y}$. Note that $\alpha$
is not separable in $x$ and $y$ since its successive derivatives $D_x^i(\alpha) = (-1)^i \prod_{j=0}^i(j+1/2)(x+y)^{-(i+3/2)}$
are linearly independent over $\bK(x)$.
Since $\beta$ is not in $\overline{\bK(x)}$. So $f$ has no telescoper of type~$(D_x, \Theta_y, D_z)$ by Theorem~\ref{THM:criteriaDSD}.
\end{example}

\section{Conclusion}

In this paper, we present existence criteria for telescopers for rational functions in three variables.
The criteria reduce the existence problems of telescopers for the trivariate inputs  to that for the bivariate inputs and two related solvable problems: the ($q$-)shift equivalence testing problem and the separation problem.
In the pure differential case,  algorithms for constructing minimal telescopers for rational functions in three variables have been presented in~\cite{Chen2012, BLS2013} using residues and reductions. This has also recently been extended to the pure shift case in~\cite{CHHLW2019} based on the existence criteria given in~\cite{Chen2016}. The first natural direction for future work is to develope efficient algorithms for other twelve cases using the existence criteria in this paper.  The next more challenging direction is to study the existence problem of telescopers for more general inputs, such as rational functions and hypergeometric terms in several variables.  To this end, we need first solve the multivariate summability problem for those inputs. In particular, it is already quite intriguing to extend the classical Gosper algorithm for indefinite hypergeometric summation~\cite{Gosper1978} to the bivariate case.

%

\bibliographystyle{plain}

\end{document}